\documentclass[11pt]{article}

\usepackage{amssymb}
\usepackage{amsmath}
\usepackage{graphicx}
\usepackage{subfig}
\usepackage[misc,geometry]{ifsym}
\usepackage{url}
\usepackage{hyperref}
\usepackage{lineno}
\usepackage{verbatim}
\usepackage{algorithm}
\usepackage{algpseudocode}
\usepackage{booktabs}
\usepackage{multirow}
\usepackage{balance}
\usepackage{verbatim}

\usepackage{setspace}
\newcommand{\Sol}[0]{{\cal X}}
\newcommand{\Neigh}[0]{{\cal N}}

\begin{document}

\title{A Local Optima Network View of Real Function Fitness Landscapes}
\author{Marco Tomassini\\ 
Information Systems Department, University of Lausanne, Switzerland}

\date{}
\maketitle

\begin{abstract}
\noindent The local optima network model has proved useful in the past in connection with 
combinatorial optimization problems. Here we examine its extension to the real continuous function
domain. Through a sampling process, the model builds a weighted directed graph which captures
the function's minima basin structure and its interconnection and which can be easily manipulated with
the help of complex networks metrics. We show that the model provides a complementary view
of function spaces that is easier to analyze and visualize, especially at higher dimension. In particular,
we show that function hardness as represented by algorithm performance, is strongly related to several
graph properties of the corresponding local optima network, opening the way for a classification of
problem difficulty according to the corresponding graph structure and with possible extensions in
the design of better metaheuristic approaches.
\end{abstract}

\section{Introduction}
\label{intro} 

Search and optimization problems arise in many important fields of science and engineering. Many of these problems
are comparatively computationally easy and enjoy efficient algorithmic solutions such as shortest paths and flows in networks, or
convex function optimization problems, just to name a few. However, a large part of the interesting
problems do not admit quick and easy solutions. This is the case, for example, of hard combinatorial optimization
problems in which the only way to find the globally optimal solution is to essentially enumerate all the admissible solutions
and to choose the best one; notable discrete examples among many others being the traveling salesman and the quadratic assignement
problems, and the global optimization of an arbitrary real function in the continuous domain. Notable improvements are possible through 
techniques such as branch-and-bound and similar complete methods or stochastic global optimization
approaches, see e.g.,~\cite{liberti} but the task remains hard.
As a consequence, many approximate techniques have been developed for both discrete and continuous hard search problems and we shall deal only with this approach in what follows.
In general, these methods
cannot offer global optimality guarantees or even certified lower bounds on the solution quality but, in exchange, 
usually they do allow finding
good enough solutions in reasonable time. The field of \textit{metaheuristics}~\cite{talbi2009} has arisen in this context and comprises
a number of approaches of the latter type; simulated annealing, evolutionary algorithms, and particle swarm optimization being well known examples. 
In contrast with classical techniques, metaheuristics and similar methods search through the space of admissible
solutions in ways that try to incrementally improve the current solution or solutions until a predefined stopping
condition is met. For this reason, having a good representation of the search space, also called the \textit{fitness landscape},
is a fundamental aspect in metaheuristic search see, for instance,~\cite{landscapes-book}. Thus, fitness landscape analysis has become an important part of
metaheuristics because the structure of the fitness landscape of a problem instance is known to be
correlated with the performance of a given search algorithm. In this context, a particular representation of a given fitness
landscape called \textit{local optima network} (LON) has been proposed recently and its usefulness demonstrated in many 
studies, see for instance,~\cite{pre08,tec11} for $NK$-landscapes,~\cite{daolio2011communities} for the quadratic assignement problem, ~\cite{hernando2017local} for the flowshop scheduling problem,
~\cite{labsTom2021} for the minimal autocorrelation binary sequences problem, and~\cite{verel2013multi} for
the application of LONs to multi-objective combinatorial optimization problems.
The LON is a graph in which the vertices are the local optima of a given problem instance and the edges
represent possible transitions between pairs of local optima. This graph thus provides a compressed representation
of the instance fitness landscape and can be analyzed in various ways. 
Once LONs are available for instances of
a problem class, the information they provide can be exploited for improving heuristic searches.
LONs were inspired by previous work in chemical physics on the structure of search spaces for atomic clusters and
of the energy landscape of protein folding~\cite{walesBook}, where the ensemble of all local optima of the energy function and their
transitions were originally called \textit{inherent structures}. Although the LON approach has mainly
be applied in the combinatorial optimization domain until recently, it can also be used in the continuous function optimization field
with some modifications. Previous research work in this direction has been presented
in~\cite{kucharik2014basin,vinko2017basin,contreras2020synthetic}. Reference~\cite{kucharik2014basin} is the
one in which, to our knowledge, basin hopping was first used for sampling the energy surface minima of
the RNA folding landscape
in order to obtain what they called ``basin hopping graph'' which is essentially a LON. Reference~\cite{contreras2020synthetic} is more closely related to the present work but it adopts a different
sampling method which leads to the discovery of ``funnels'' of minima, to be explained later, instead of
the full LON.
Reference~\cite{vinko2017basin} is
particularly relevant for the present work as we use essentially the same methods and points of view in the
LON construction and analysis. However, we extend~\cite{vinko2017basin} in several ways, especially in the direction
of scaling up the number of space dimensions, in examining the role of edge weights, and in relating complex networks metrics with the empirical hardness of the tested functions.

Following those previous studies, in this paper we apply the LON idea to the space of continuous real-valued functions.
After introducing the LON concept, we describe the sampling methodology which is different from the one used in discrete 
fitness landscapes and is based on an optimization heuristic called \textit{basin hopping}, to be explained later. 
This is followed by a study of the LONs generated from a few well known highly multimodal test functions
with the help of complex networks tools and metrics. We then study the relationship between these metrics
and the empirical hardness of functions and we extend the analysis to higher dimension. We conclude with
a discussion on the pros and cons of the network approach and with some indications of interesting current
and future work.

\section{Search Spaces and Local Optima Networks}
\label{lons}

In this section we summarize the customary LON model for combinatorial optimization problems, followed by
a description of the changes needed for dealing with real function global optimization.

\subsection{Local Optima Networks in Discrete Spaces}

Given a discrete problem $ \mathcal P$ and an instance $\mathcal I_{\mathcal P}$ of $ \mathcal P$, 
$I_{\mathcal P}$'s \textit{fitness landscape}~\cite{reidys2002combinatorial} is a triplet $(\Sol, \Neigh, f)$ with
 $\Sol$ is a set of admissible solutions; $\Neigh : \Sol \longrightarrow 2^\Sol$, a function
that assigns to every $x \in \Sol$ a set of neighbors $\Neigh(x)$, and $f : \Sol \longrightarrow \mathbb{R}$ 
gives the fitness or objective value  $\forall x \in \Sol$. The neighborhood $\Neigh(x)$ of a solution $x$
depends on the ``move'' operator used to produce a new solution from the current one and on the
solution representation. For example, if solutions are expressed 
as bit strings $x$ of length $k$, the mutation of one bit
switches a random bit of $x$ and so $\Neigh(x)$ is the set of the $k$ strings at distance one from $x$.
Different neighbohoods can be defined depending on the move operator that has been selected. 

The LON concept starts from the fitness landscape and compresses it by building a weighted 
directed graph $G_w=(V,E)$ in which the set of vertices $V$ are the
\textit{local optima} in the search space and $E$ is the set of directed edges joining
two vertices (a.k.a. optima) when they are directly connected.
Here we assume minimization, maximization is the same with the obvious changes.
For a minimization problem a solution $x \in \Sol$ is a local optimum iff $\forall x^{\prime} \in \Neigh(x)$, $f(x^{\prime}) \geqslant f(x)$.

The local minima, i.e., the vertices $V$, are obtained using a best-improvement descent local search in an exhaustive way. Thus, each 
admissible solution $x$ is assigned to a given local optimum, the one at which the local search stops.
At the end of the process the whole search space $\Sol$ becomes partitioned into a number $N$ of basins of attraction $B_i$,
$\Sol = B_1 \cup B_2 \cup \ldots \cup B_N$, with $B_i \cap B_k = \emptyset, \forall i \ne k$.

The edges $E$ of the network can be defined in different ways which are almost equivalent, for details see,
e.g.,~\cite{pre08,labsTom2021}. In all cases, these directed edges stand for possible transitions between the
local optima that they join. The frequency of transition is given by the edge's weight.
The exhaustive extraction of the problem's LON is possible for small to medium size problem instances. Larger instances can also be tackled using the sampling methodology of~\cite{thomson2019}.

\subsection{LON Methodology for Real Function Optimization}
\label{real-lons}

We remark that ideas essentially similar to the ones described above for combinatorial spaces 
 were already present in a 2005 paper by M. Locatelli~\cite{locatelli2005multilevel} in which he discusses the multilevel
 structure of global optimization problems in the continuous domain. 
Now we discuss the changes needed in the above description to be able to deal with the search spaces of real functions
in $\mathbb{R}^n$. Given a function $f$,
the global minimization problem is to find the minimum value $m$
 of $f$ over a feasible domain \raisebox{2pt}{$\chi$} and can be stated as follows:

\begin{equation*}
\begin{aligned}
&\underset{\textbf{x}}{\text{min}}
& & \hspace{-0.3cm} \{f(\textbf{x}) \;:\; \textbf{x} \in \raisebox{2pt}{$\chi$}\}  \\
%& \text{subject to}
%& & \textbf{x} \in \raisebox{2pt}{$\chi$}
\end{aligned}
\end{equation*}

\noindent Usually, one also wants to know the argument, i.e., the 
point, or set of points, $\textbf{x}$ that provide the minimum value $m$ of the function:

\begin{equation*}
\begin{aligned}
& \underset{\textbf{x}}{\text{argmin}}
& &  \hspace{-0.3cm} \{\textbf{x} \in \raisebox{2pt}{$\chi$} \; : \; f(\textbf{x}) = m \}  \\
\end{aligned}
\end{equation*}

\noindent Here $\textbf{x}$ is a real column vector of scalar variables $[x_1, x_2, \ldots, x_n]^T, \;\;\forall x_i \in \mathbb{R}$. 
Maximization is obtained by replacing $f(\textbf{x})$ with $-f(\textbf{x})$. The definition also
covers constrained optimization problems with a proper definition of $\raisebox{2pt}{$\chi$}$. However, here
we will only consider so-called ``box constraints'' that limit each variable to a segment of the real line, thus restricting the
search space to the hyperrectangle $\chi = \{\textbf{x} : \textbf{l} \le \textbf{x} \le \textbf{u}\}$, where $\textbf{l}$
and $\textbf{u}$ are vectors of size $n$ defining the respective lower and upper bounds on each coordinate of \textbf{x}.

In contrast with the discrete case, in topology the neighborhood of a given point $\textbf{p}$ in a real, continuous metric space is defined as an open ``ball'' of a given radius $r$ around $\textbf{p}$, which is
the set of all points in $\chi$ that are at distance $d$ less than $r$ from $\textbf{p}$. Here the standard
Euclidean distance is assumed.

In practice, the ball radius will depend on the function of interest and should be chosen appropriately in each
case. Having defined the neighborhood, we now need to say how the vertices and edges of the LON $G_w=(V,E)$ are obtained.
A method that is capable
of finding the minima of a function, as well as the transitions between them called \textit{basin hopping} (BH), is used as a sampling technique.

The outline of the BH algorithm is very simple, see pseudocode~\ref{bh}, where solutions
$s,x,y,z$ are to be understood as $n$-dimensional vectors.
\begin{algorithm}
\caption{Basin Hopping}
\label{bh}
\begin{algorithmic}[0]
\State $s \leftarrow$ generate initial solution
\State $x \leftarrow $ minimize($f(s)$)
\While{termination condition not met}
\State $y \leftarrow $ perturb($x$)
\State $z \leftarrow $ minimize($f(y)$)
%\IF{$f(y) < f(x)$}
\State$x \leftarrow $ acceptance(x, z)
%\STATE $x \leftarrow y$
%\ENDIF
\EndWhile
\State \textbf{return} $x, f(x)$
\end{algorithmic}
\end{algorithm}
\noindent   The algorithm starts by generating an initial solution $s$ either randomly or heuristically. Unless the fitness landscape is flat, this current
solution must belong to the basin of attraction of some local optimum whose coordinates $x$ are found
by using a local search procedure starting at $s$. After that the algorithm iterates  three
stages. First, the current solution $x$ is perturbed by some kind of coordinates change yielding the new solution $y$. 
Next, starting at $y$, a local minimizer finds the new local minimum $z$. There are two possibilities: either $z$ is
different from $x$ or it is the same. In the first case, the algorithm has successfully jumped out of the basin of attraction
of $x$. Otherwise, the perturbation has been insufficient and the point $y$ belongs to the original basin of attraction,
causing the search to find the same minimum again. Finally, the acceptance phase consists in deciding whether
the new solution $z$ is accepted as the starting point of the next cycle. If it is the same as before the perturbation then
it is accepted and the search resumes by trying another perturbation from this point. Otherwise, it is accepted
subject to some condition. For example, it could be accepted only if $f(z) \leqslant f(x)$, or with some probability $p$
that decreases with increasing difference $|f(z)-f(x)|$ if $f(z) > f(x)$. The search then continues
from the new minimum $z$. The BH heuristic comes from the field of chemical physics and has been successful
in finding energy-optimal structures of atomic clusters and simple proteins, see e.g.,~\cite{wales1999}.

As a sampling technique, to our knowledge BH was first used by Kucharik et al.~\cite{kucharik2014basin}
 to characterize the folding landscapes of RNA for building the
corresponding LONs which were called \textit{basin hopping graphs}. It was also used
in follow-up work on continuous functions by Vink\'o and Gelle~\cite{vinko2017basin} and by 
Contreras-Cruz et al.~\cite{contreras2020synthetic}.
Our technique for sampling the function minima and their interconnections is built around the BH skeleton but
differs from the method of~\cite{contreras2020synthetic}
in that a new local optimum is always accepted regardless of its objective function value, since the goal here is to sample
a sufficient number of optima besides the best one(s) in order to build a good approximation of the actual LON. The
algorithm is outlined in pseudocode~\ref{bh-sample} where $s,x,y$, and $z$ are to be understood as $n$-dimensional
vectors.

\begin{algorithm}
\caption{Basin Hopping Sampling}
\label{bh-sample}
\begin{algorithmic}[1]
\Require $f(x)$, $x \in \Sol, \Sol \subseteq \mathbb{R}^n$, strength $p$, number of sample points $q$
\State $V = \emptyset, E=\emptyset\;\;\;\;$  //vertices and edges sets
\For {each sampling point $s$ in $1..q$}
%\State $s  \leftarrow $ initial solution
\State $x \leftarrow $ minimize($f(s)$)
\State $V \leftarrow V \cup \{x\}$
\While{termination condition not met}
\State $y \leftarrow $ perturb($x,p$)
\State $z \leftarrow $ minimize($f(y)$)
    \If {$z  \not\in V$}
    	 \State $V \gets V \cup \{x\}$
	 \State$E \leftarrow  E \cup \{xz\} $
               \State $w_{xz} \gets 1$
           \Else  \If {$z = x$}
                \State $w_{xx} \gets w_{xx} + 1$
               \Else 
               \State $w_{xz} \gets w_{xz} + 1$
       	  \EndIf
    \EndIf
    \State $x \gets z$
\EndWhile
\EndFor
\State $\textbf{return} G_{w}(V,E)$
\end{algorithmic}
\end{algorithm}

We now explain the sampling process in more detail. To start with, a real function $f$, a search domain $\Sol \in \mathbb{R}^n $, a perturbation strength $p$, and a number $q$ of sampling points are given. The perturbation strength $p$ is important and must be set by trial and error in order to fit
the particular function $f$ hypersurface. Smaller perturbations are more useful to discover local minima in a
highly multimodal function while larger perturbations are more adapted to slowly varying functions. The number of sampling 
points $q$ depends on the dimensionality of the space $n$ but it cannot grow exponentially as the potential number of 
points to be sampled does. We simply make it proportional to $n$: $q = kn$ with $k$ a small integer. Clearly, this will 
cause the sampling to be less complete as $n$ increases but in this way the task remains feasible. The $q$ points
can be generated uniformly at random but a better way is to generate them such that they cover the region as well as possible.
Here we use quasi-random space-filling sequences obtained according to the Sobol method~\cite{press2007}. This ensures that the initial
sampling points are minimally correlated. Next, for each sampling point a \textbf{while} loop  is executed (line 5)
The termination condition of the \textbf{while} loop is set as a maximum number of allowed function evaluations.
Each time a new minimum is found\footnote{We use the standard quasi-Newton method L-BFGS from the Python SciPy library which
can also work by numerically approximating the derivatives if they are not available analytically.}, it is added to the set of graph vertices $V$ (lines 4 and 9). Each new directed edge $\{x,z\}$, with
direction from $x$ to $z$, is added to the set of edges $E$ (line 10) or, if the edge existed already, its weight is increased (lines 14 and 16) and the search continues from the last minimum found. Notice that self-loops simply count the number of times a perturbation followed by local minimization fell back into the
basin of attraction of the minimum from which the perturbation was applied. In continuous spaces there is a numerical 
precision issue that is not present in discrete spaces: when the search finds an optimum with the same function value as a
previous one (to a certain precision), one must decide whether the optimum has already been found or 
if it is new (lines 8 and 12). 
The problem is easily solved by a test on the optima coordinates: the absolute value of their differences must be
less than $\epsilon$ for all the coordinates for the optima to be the same. We have used $\epsilon = 0.0001$.
At the end the LON $G_{w}(V,E)$ is returned.
The above sampling procedure differs from the method used
in~\cite{contreras2020synthetic}. In the latter, only minima that are monotonously decreasing in objective value, i.e., only
improving or equal solutions are accepted. As a consequence, the resulting sampled graphs have long linear branches.

\begin{figure*}[]
  \begin{center}
   \includegraphics[width=0.6\textwidth]{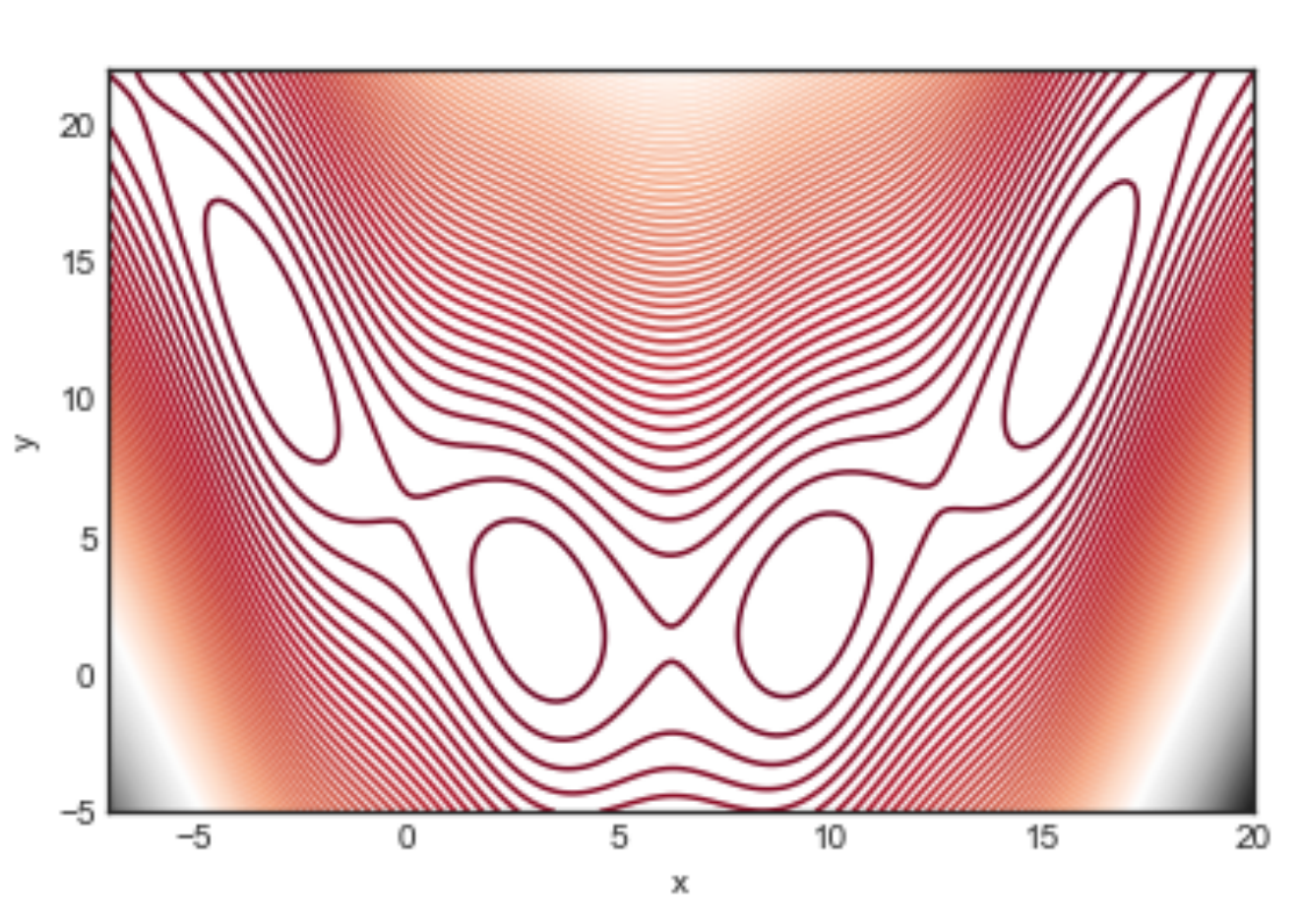}  
   \end{center}
  \caption{Contour lines representation of the Branin function $B(x,y)$. In the given domain the function has four 
 global minima named $0,1,2,3$ from left to right and located at $(-\pi,12.275),(\pi,2.275),(3\pi,2.475),(5\pi,12.875)$ with $B(x,y) \approx 0.39789$.}
\label{branin}
\end{figure*}

\begin{figure*}[]
  \begin{center}
   \includegraphics[width=0.5\textwidth]{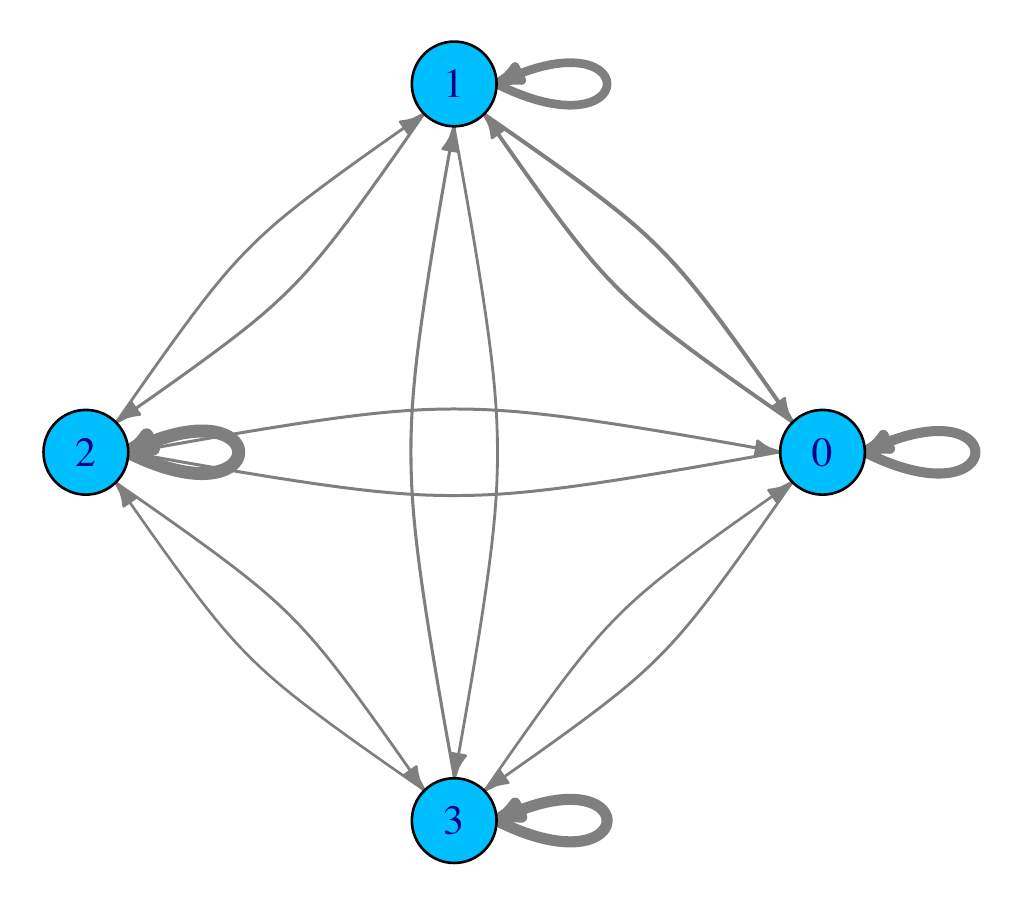}  
   \end{center}
  \caption{The LON corresponding to the function Branin shown in Fig.~\ref{branin}. Vertex labels correspond to the
  optima and edge thickness is 
  drawn proportional to the corresponding edge weight, which stands for the empirical frequency
  of transition between the corresponding minima basins.}
\label{braninlon}
\end{figure*}

A simple example should be useful to illustrate the above procedure. Let us consider the Branin function $B(x,y)$ in two dimensions
which is defined as follows:

$$
B(x,y) = \bigg(-1.275\,\frac{x^2}{\pi^2} + 5\,\frac{x}{\pi} + y - 6\bigg)^2 + \bigg(10 - \frac{5}{4\pi} \bigg) \cos(x) + 10
$$

\vspace{0.05cm}

The function is depicted in Fig.~\ref{branin} by means of contour lines. All of its minima are global with
$x = \pi + 2\pi k$ with $k$ a positive or negative integer. In the region $[(-7,18),(-5,20)]$ shown in the
figure there are four global minima at: $\{(-\pi,12.275),(\pi,2.275),(3\pi,2.475),\\(5\pi,12.875)\}$ with
a function value $B(x,y) \approx 0.39789$.

For such a small number of optima and limited search region, the sampling can be exhaustive, in the sense
that all the minima are found and all the possible transitions are recorded too. Figure~\ref{braninlon} shows
the corresponding LON. Actual edge weights are not shown to avoid cluttering the figure but the edge thickness is drawn
proportional to them. What can be said is that many transitions
fall back into the same starting basin and thus the weights of the self-loops are generally higher. Also, by numbering
the minima $0,1,2$, and $3$ from left to right in Fig.~\ref{branin}, what the edge weights of the LON say is that
transitions to and from the central adjacent minima $1 \rightarrow 2$ and $2\rightarrow1$ are more frequent than transitions between $0$ and $1$ and between
$2$ and $3$ in both directions. Transitions between more distant minima, i.e., between $0$ and $2$, $1$ and
$3$, and $0$ and $3$ are less frequent, as expected. We end this section with a comment on \textit{neutrality}.
Neutral fitness landscapes are those in which there exist large regions in which the objective function
has the same value. This notion is particularly important is discrete problems such as the Satisfiability Problem (SAT) and many
others. However, most mathematical functions of interest for optimization are either monotone increasing or decreasing,
except at special points and thus neutrality is normally not an issue. At the LON level, however, there can be some neutrality.
For instance, in the Branin function above the four degenerate global minima are a neutral set. Neutrality can be
easily dealt with in LONs by compressing all connected nodes at the same fitness into a single
representative node as explained in~\cite{tec11}. Thus a compressed LON for the Branin function would consist of a single
node. However, doing so would deprive us of the topological knowledge implied by the use of the whole neighborhood
of a given optimum which has important implications for search techniques.
For this reason, and also because neutrality is usually scarce in LONs derived from real-valued function spaces, we shall 
not compress neighbors nodes in the sequel and shall represent all of them in the sampled LONs.

\section{LONS of Some Common Test Functions}
\label{functs}

The goal of this section is to apply the previous concepts to typical functions belonging to certain
classes.
 When testing new global optimization algorithms it is customary to evaluate them on a given set of 
 contrived functions that are chosen to represent the main kinds of features found in functions arising
 from general real-life optimization problems. No amount of testing of this kind can prove that an algorithm, say $A_1$,
 is definitely better than another algorithm $A_2$ on all possible problems and problem instances.
 However, such benchmarking is widely used and good results on a sufficiently complete function test set 
 may provide at least some confidence on the expected behavior of an algorithm in general cases.
 For a detailed presentation of benchmarking methodology and a discussion of its limitations see,
 e.g.,~\cite{bartz2020bench}.
 
There exist hundreds of test functions in the literature. Some previous work on synthetic real-valued functions making use of the 
LON representation has appeared in~\cite{vinko2017basin,contreras2020synthetic}. Studying many 
of them would be too time-consuming but
their number can be usefully limited by selecting only a few, provided that they contain the principal structural
features that play a role in global optimization. 
We shall concentrate on some of those functions devoting attention to the relationships between
some general properties of function structure and their LON. Among the most important properties, 
we consider \textit{separability/non-separability}, \textit{unimodality/multimodality}, and their combinations.
To start with, unimodal functions give rise to a LON with a single node. There is almost nothing to be learned from such
a graph and thus we ignore this type of functions in the sequel.
We now consider two typical scalable representative test functions of the highly multimodal type: the Rastrigin function and the
Griewank function whose expressions are given below. The main structural difference is that Griewank is non-separable while Rastrigin is separable. In short, separability means that each coordinate can be optimized independently of the others
while this is not possible in non-separable functions in which variables interact. In practice this means that non-separable problems are often harder to solve. We start with the two-dimensional versions of these functions as they are easier to
understand and visualize. The effect of scaling up the dimension $n$ will be investigated later on.
%It is standard practice to use rotation matrices in order to turn a separable function into a non-separable one. 

\paragraph{\textbf{Rastrigin function:}}

$$
R(\textbf{x}) = 10\, n +  \sum_{i=1}^n [x_i^2 - 10\, \cos(2 \pi x_i)]
$$

Search region:  $\textbf{x} \in[-30,30]^n$. Global minimum at $ \textbf{x}^* = (0, \ldots, 0)$ and $R( \textbf{x}^*) = 0$.\\

\paragraph{\textbf{Griewank function:}}

$$
G(\textbf{x}) = \sum_{i=1}^n \frac{x_i^2}{4000} - \prod \cos \left( \frac{x_i}{\sqrt i} \right) + 1
$$

Search region: $\textbf{x} \in[-60,60]^n$. Global minimum at $ \textbf{x}^* = (0, \ldots, 0)$ and $G( \textbf{x}^*) = 0$.\\

\begin{figure*}[h!]
  \begin{center}
   \includegraphics[width=0.45\textwidth]{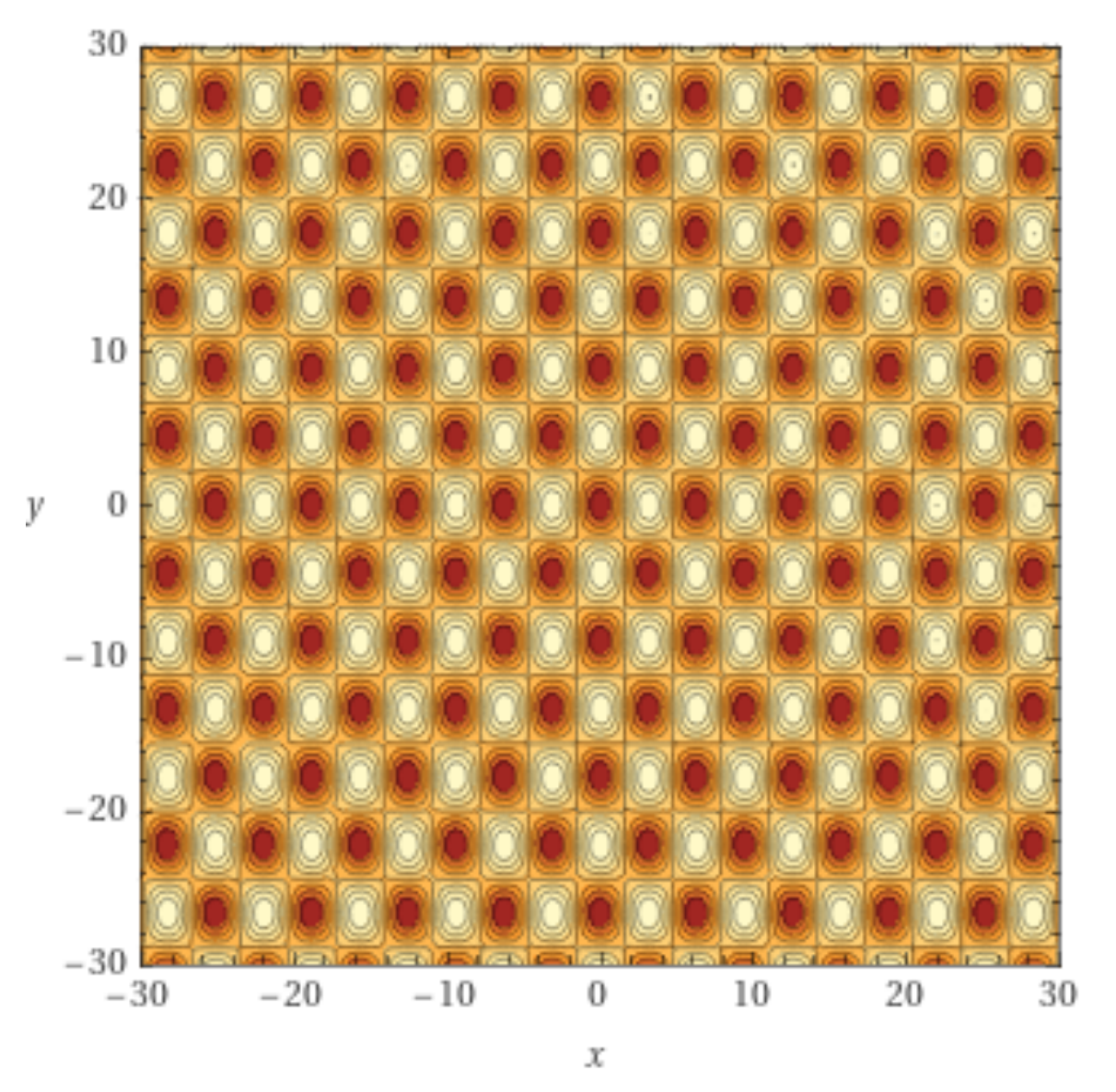}  
    \includegraphics[width=0.45\textwidth]{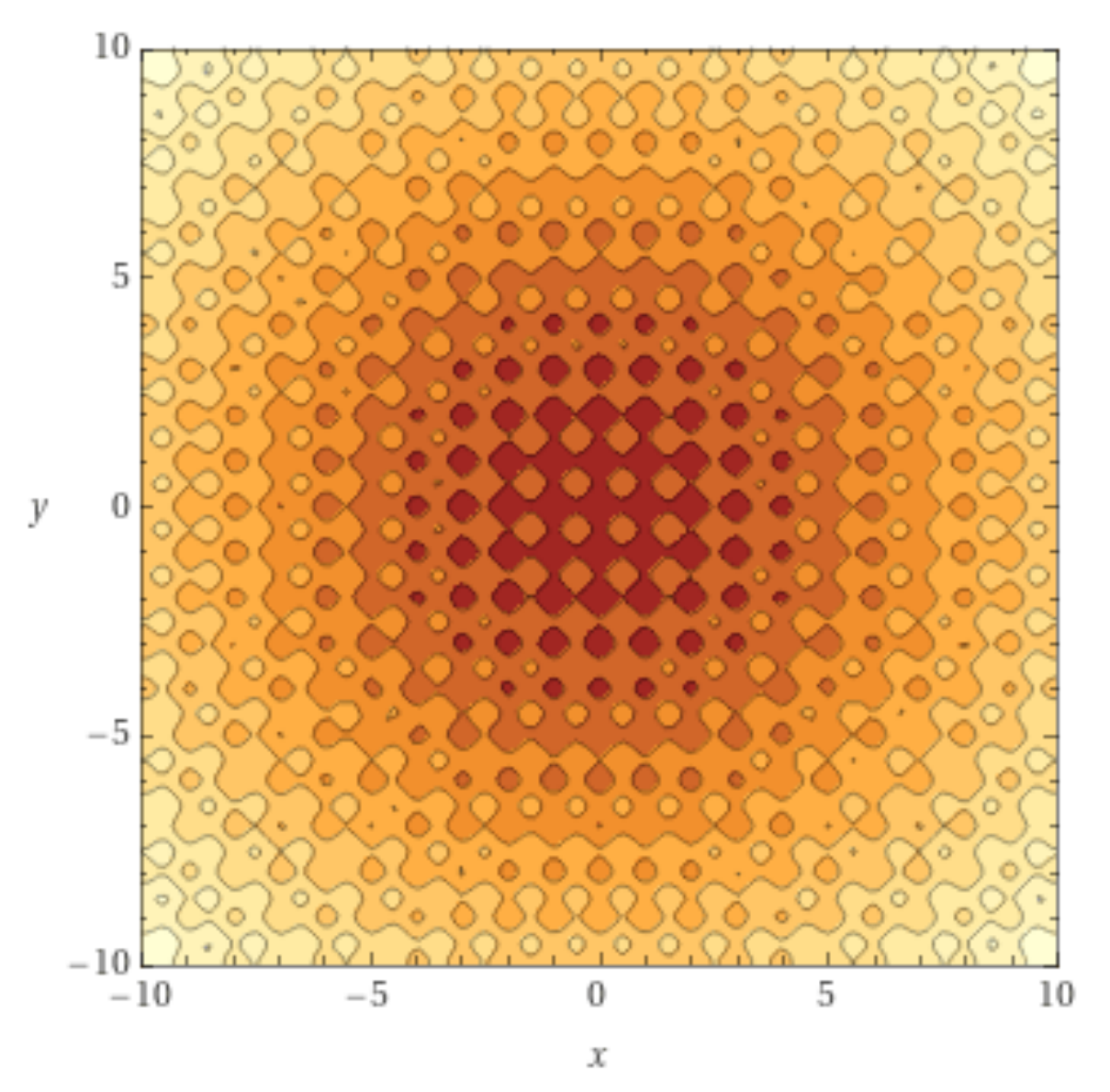}  
      \end{center}
  \caption{Contour lines plots for two-dimensional Griewank (left) and Rastrigin (right) functions.
  Darker areas contain the local minima. Note the different region sizes.}
\label{gw-cont}
\end{figure*}

Figure~\ref{gw-cont} shows the two-dimensional versions of functions $R(\textbf{x})$ (right image) and
$G(\textbf{x})$ (left image). Both are highly multimodal, with symmetrical equal-valued minima 
around the global optimum at $(0,0)$. The Rastrigin function has a higher density of
local optima (notice the different region sizes).

\noindent We shall also use the Schwefel function, defined as follows:

\paragraph{\textbf{Schwefel function:}}
$$
S(\textbf{x}) = n*418.9829 - \sum_{i=1}^n x_i \sin(\sqrt{( \lvert x_i\rvert)}
$$

\begin{figure*}[h!]
  \begin{center}
   \includegraphics[width=0.45\textwidth]{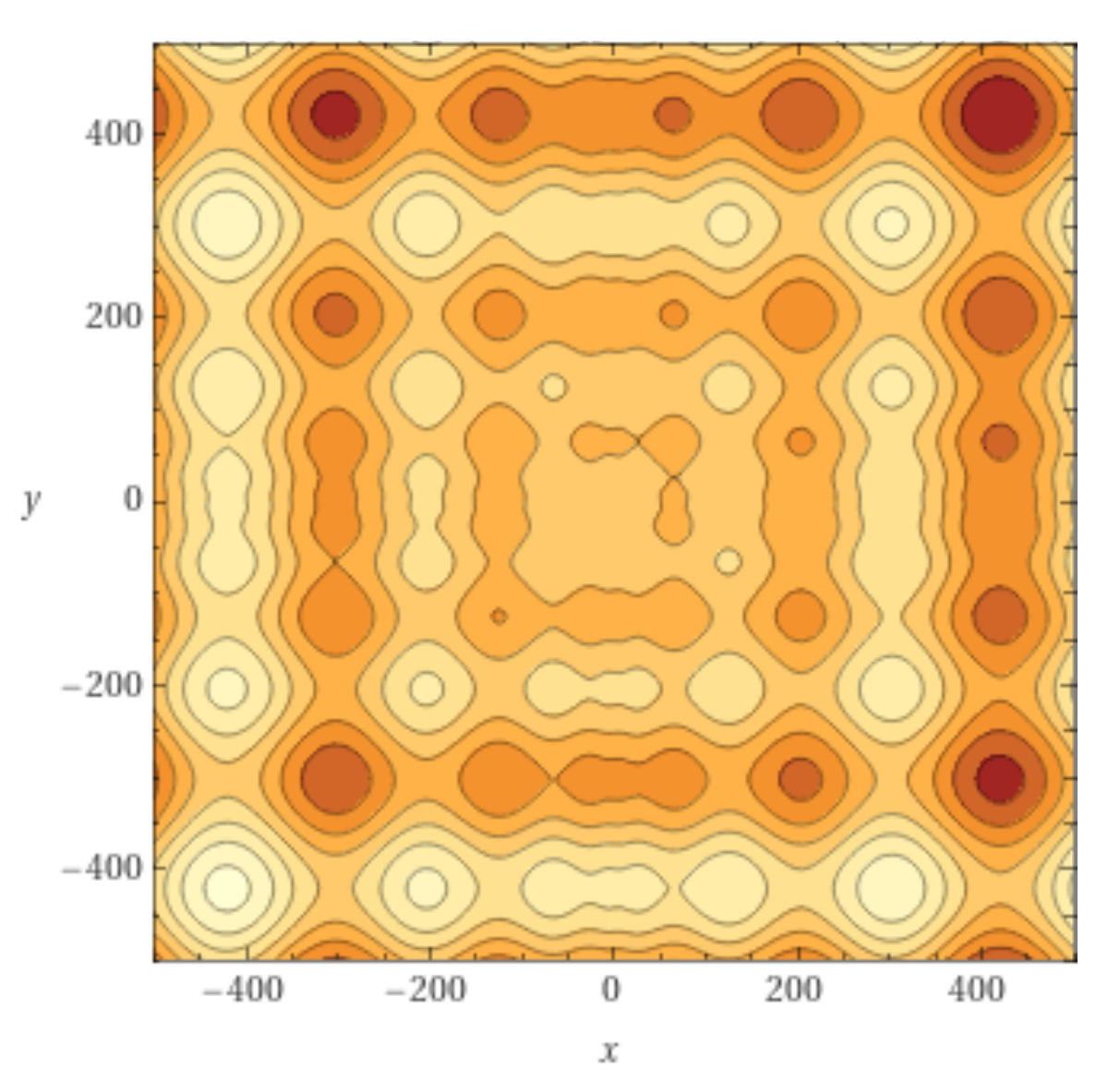}    
      \end{center}
  \caption{Contour lines representation of the two-dimensional Schwefel function in the domain
  $x \in[-500,500], y\in[-500,500]$.}
\label{schwefel}
\end{figure*}

Search region:  $\textbf{x} \in[-500,500]^n$. Global minimum at $ \textbf{x}^* = (420.968746, \ldots, 420.968746)$ and $S( \textbf{x}^*) = 0$.
The contour lines structure of the two-dimensional $S$ function is shown in Fig.~\ref{schwefel}.

\subsection{Algorithm Performance and Problem Hardness}
\label{perfs}

Some LON metrics are related to the hardness of the function being optimized. It is thus necessary 
to obtain an empirical evaluation of the optimization hardness of each function and we do this here
for the Griewank and Rastrigin functions. Since there is no agreed upon measure of hardness for
arbitrary real functions, we follow common usage and take as a reasonable empirical
proxy the success rate over a large number of optimization runs. Due to limited numerical precision, a run
is considered successful when the final function value is within $0.0001$ of the known global minimum.
To reduce algorithm biases, we used two different metaheuristics: differential evolution (DE)~\cite{storn1997} and basin hopping \footnote{both
algorithms were taken from the Python SciPy optimization library using the library standard parameter settings.}.

\begin{table}[h!]
\small
\begin{center}
\vspace{0.3cm}
\begin{tabular}{|l|c|c|}
\hline
\multicolumn{3} {|c|}{Success Rates}  \\
\hline
 & DE & BH   \\
\hline
Griewank &  \;\; $0.464\,(2479)$  \;\;  &  \;\;\ $0.050\,(1528)$  \;\;   \\
\hline
Rastrigin &  \;\;    $0.930\,(2178)$ \;\;  &  \;\; $0.980\,(928)$  \;\;    \\
\hline
\end{tabular}
\vspace{0.3cm}
\caption{Fraction of optimization runs that found the global minimum for the Griewank and Rastrigin
functions in two dimensions for DE (left column) and BH (right column).
The average is over $500$ optimization runs in each case. 
A budget of $3.0 \times 10^3$ function evaluations has been allowed per run and the search region is
$[-30,30]^2$. The mean number of function evaluations
when the global minimum has been found is in parentheses.}
\label{hard}
\end{center}
\end{table}

We run each algorithm $500$ times on each function and recorded the 
number of hits and the average number of function evaluations for successful executions. The results are in Table~\ref{hard}. We point out that these results, as well as those in Table~\ref{D5-10} are well known; they have been replicated here to make them readily available for the sake of the reader.
It is clearly seen that, at least for these functions and the choice of algorithms parameters, BH outperforms DE 
in terms of  efficiency (average number of function evaluations) but DE, probably
because of its better diversification features due to a population of searchers is more successful on
Griewank. However, here we are
mainly interested in the ranking and in both cases Griewank turns out to be the hardest of the two functions.

\begin{table}[h!]
\begin{center}
\vspace{0.3cm}
\footnotesize
\begin{tabular}{|l|c|c|c|c|}
\hline
\multicolumn{5} {|c|}{Success Rates}   \\
\hline
% aaa & bbb& \multicolumn{2}{c|}{fusion1} & \\
& \multicolumn{2}{c|}{n = 5} & \multicolumn{2}{c|}{n = 10} \\
\hline
 & $DE$ & $BH$ &  $DE$  & $BH$\\
\hline
Griewank &  \;\; $0.157\,(25496)$  \;\;  &  \;\;\ $0.080\,(18575)$  \;\;& \;\; $0.016(96573)$ \;\;&  \;\;$0.926(75314)$ \;\;    \\
\hline
Rastrigin &  \;\;    $0.818\,(14754)$ \;\;  &  \;\; $0.352\,(14945)$ \;\;  &  \;\;\; $0.496(105202)$  \;\; &  \;\; $0.165(107571)$ \;\;     \\
\hline
\end{tabular}
\vspace{0.3cm}
\caption{Fraction of optimization runs that found the global minimum for the Griewank and Rastrigin
functions for dimension $n=5$ (left half of the table) and $n=10$ (right half) using DE and BH.
The averages are over $500$ optimization runs in each case. 
A budget of $3.0 \times 10^4$ function evaluations has been allocated per run and the search region are,
respectively, $[-30,30]^5$ and $[-30,30]^{10}$. The mean number of function evaluations
when the global minimum has been found is in parentheses. }
\label{D5-10}
\end{center}
\end{table}

These results are confirmed when scaling the functions to higher dimensions except in the Griewank case. Table~\ref{D5-10} gives the results for dimensions $n=5$ and $n=10$. The overall view is that Rastrigin 
becomes progressively harder with increasing dimension but remains easier than Griewank. The exception
is Griewank with $n=10$ for which the function is hard using DE but becomes easy with BH. This apparent
``anomaly" has a straightforward explanation given by Locatelli in~\cite{locatelli2003}. Locatelli found that the Griewank function $G(\textbf{x})$ becomes
easier with increasing dimension by using a naive multistart algorithm. In fact, $G(\textbf{x})$
is the sum of a quadratic convex part and an oscillatory non-convex part and Locatelli showd that, since
the convex part minimum is also the global minimum, as $D$ increases the gradient of the non-convex
part becomes smaller and smaller and the search becomes more and more
dominated by the convex minimization which is easy for algorithms of the Quasi-Newton type 
like BFGS~\cite{broyden1967}. Now, BH is structurally similar to the naive multistart as it uses a BFGS-type
routine for local minimization, the only important difference being that start points for local minimization
are not randomly chosen in the search space; instead, they are found by jumping in the neighborhood of the
last minimum found. BH thus follows a trajectory through the function minima but is otherwise similar
to the multistart approach, which explains its efficiency at higher dimensions. On the other hand, DE is 
derivative-free and thus it must deal with the exponentially increasing number of local minima without the
help of a numerical minimizer as the problem dimension increases.

\subsection{Local Optima Networks Statistics}
\label{metrics}

The LON methodology is useful mainly for moderately to
highly multimodal real functions. Fortunately, these are often the type of functions that are hard
to minimize globally and are often found in real world applications too.
Apart from small LONs like the one corresponding to the Branin function, direct visualization of the graph
quickly becomes impractical and of little help as the number of nodes increases beyond a few hundreds.
Instead, our methodology is based on complex networks concepts and consists in computing
a number of network measures that are known to correlate with problem
difficulty and that have been previously found useful in the study of the LONs of problems defined in
combinatorial spaces~\cite{pre08,tec11,daolio2011communities,hernando2017local}. Based on previous
experience, here we use the network metrics briefly defined below for the sake of self-containedness.
Fuller explanations can be found in a textbook such as~\cite{newman2018}.

\paragraph{\bf Node strength.} 
The strength $s_i$  is the generalization of the vertex degree to weighted networks and 
it is defined as: 

\begin{equation}
s_i = \sum_{j \in \Neigh(i)} w_{ij},
\end{equation}

\noindent where $w_{ij}$ is the weight of the edge $(i,j)$ and the sum is over
all the neighbors $\Neigh(i)$ of a node. In directed graphs, like those we study here,
we have, respectively, the outgoing strength which is computed summing the weights of the outgoing edges
from a node,  and
the incoming strength computed analogously from the incoming edges to a node. Self-loops may
be counted either as incoming or outgoing but not both.

\paragraph{\bf Degree distribution functions.}
The degree distribution $P(k)$ provides, for each degree $k$, the number of nodes having this degree
or its frequency. For directed graphs there are two distinct degree distributions: one for the incoming
edges and another for for the outgoing edges. These distributions, and their expected values, are very
useful for characterizing the graph as being homogeneous or heterogeneous in terms of the degrees
of its vertices.

\paragraph{\bf Mean shortest paths.} 

The average value $\langle L \rangle $ of the shortest distances $l_{ij}$ for all the possible pair of vertices
$i,j$ in a network is given by:

\begin{equation}
\langle L \rangle =   \frac{2}{N(N-1)}\sum _{i=1}^N \sum _{j > i} l_{ij},
\label{av-pl}
\end{equation}

\noindent 
Where $N=|V|$ is the number of vertices. Paths are defined only for connected graphs. Since 
LONs are directed, we require in addition that they be strongly connected. 
The longest among all the shortest paths is called the \textit{diameter} of the graph.

\paragraph{\bf Centrality.}
The importance of a vertex in a network can be assessed with the help of \textit{centrality} measures.
There are several such measures in common use; 
here we investigate centrality of optima in a LON network with the use of PageRank.

\section{Results and Discussion}
\label{resul}

Below we present first our results of applying the chosen network metrics to the LONs of the Rastrigin and Griewank
two-dimensional functions. All computations have been done with custom software, igraph R, and Python NetworkX
\footnote{All the data used in the analysis will be made available by the author upon request.}.

\subsection{Strength}
\label{stren}

\begin{figure*}[h!]
  \begin{center}
   \includegraphics[width=0.495\textwidth]{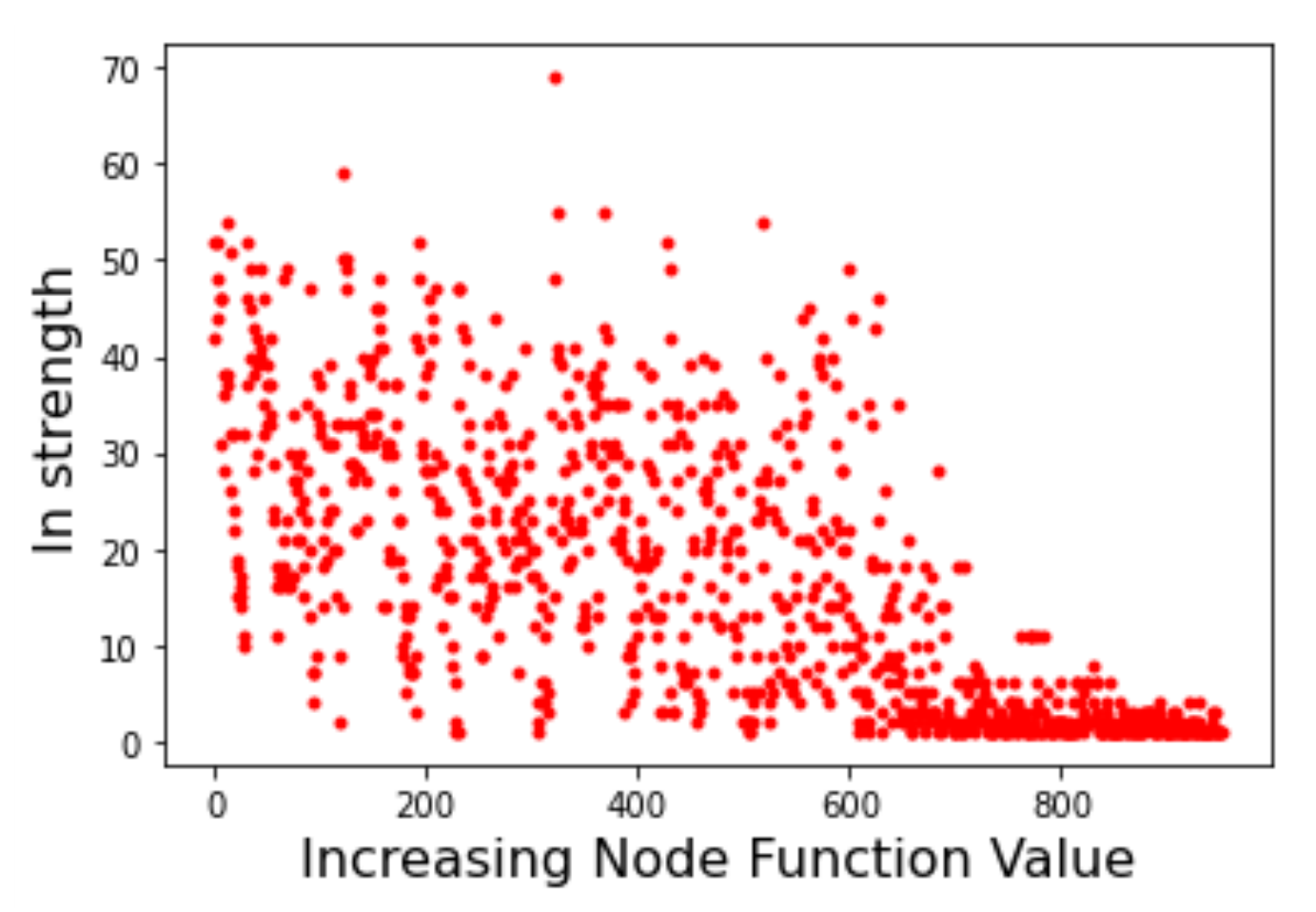}
   \includegraphics[width=0.495\textwidth]{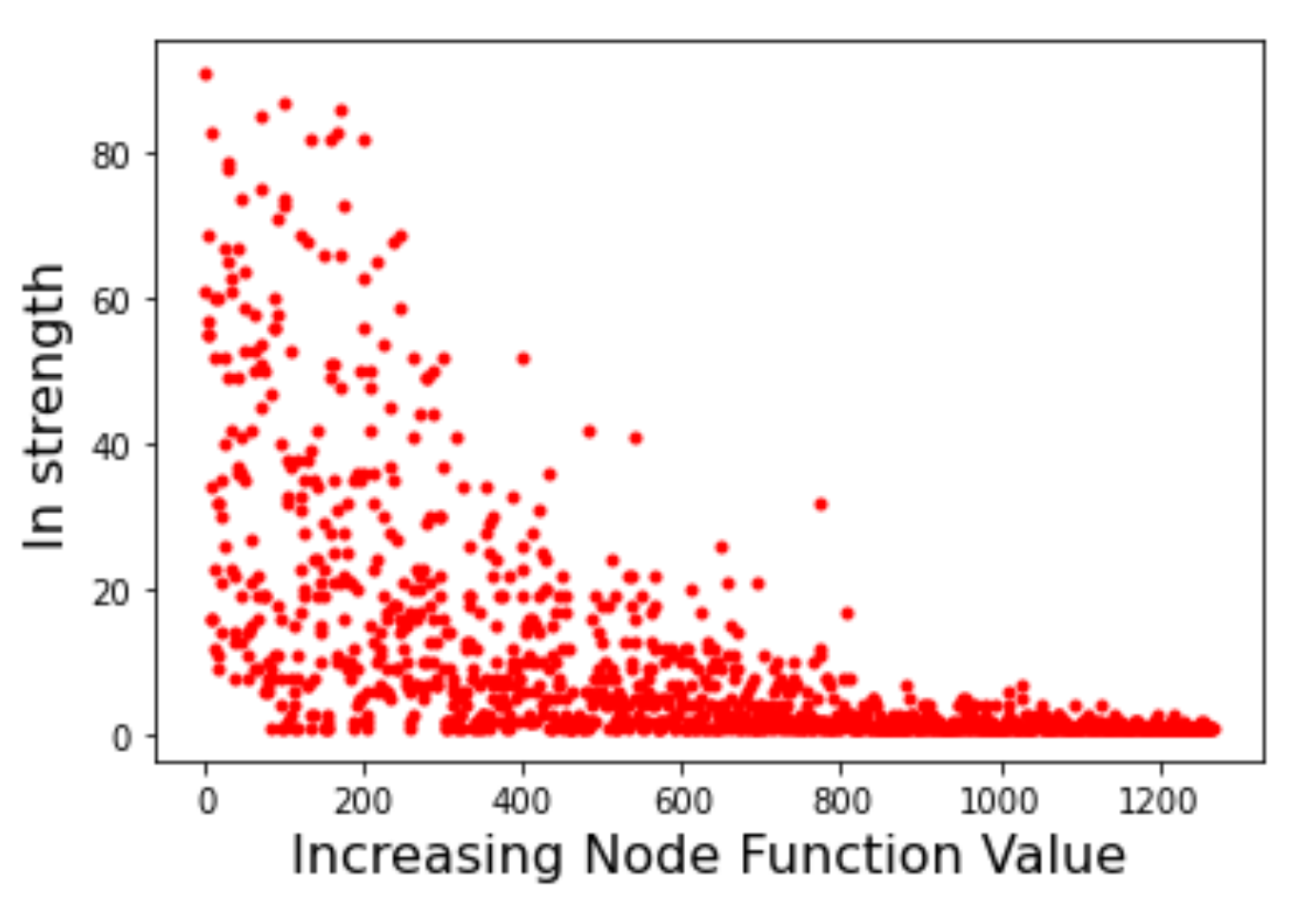} 
   \end{center}
  \caption{Incoming strength of LON nodes. Left image:  Griewank function. Right image: 
  Rastrigin function. Graph nodes are sorted by increasing objective function value.}
\label{stren}
\end{figure*}

Node strength is related to the probability of transition between
minima basins. As such, it can provide useful information on the likely behavior of 
optimization methods that search the energy landscape. Figure~\ref{stren} depicts the strength of the
incoming edges to a node for the Griewank LON (left image) and the Rastrigin LON (right image) when the
LON nodes are sorted by increasing function value of the corresponding minima to the right of the x-axis, i.e.,
the best minima are near the origin. We see that in the Griewank case there is no apparent relationship between incoming strength and minima fitness.
 On the other hand, for Rastrigin there is a clear qualitative correlation: nodes corresponding to good minima have
 high incoming strength. This can be interpreted saying that the best minima are more likely to
 be reached as they have many more incoming edges from surrounding nodes. Since nodes also represent
 basins of attraction of the corresponding minima, we can also say that transitions to high fitness basins are more likely
 in the Rastrigin case. This is in agreement with the fact stated in Sect.~\ref{perfs} that Griewank is harder to
 optimize than Rastrigin in two dimensions.

\subsection{Degree Distribution Functions}
\label{DDF}

\begin{figure*}[h!]
  \begin{center}
   \includegraphics[width=0.495\textwidth]{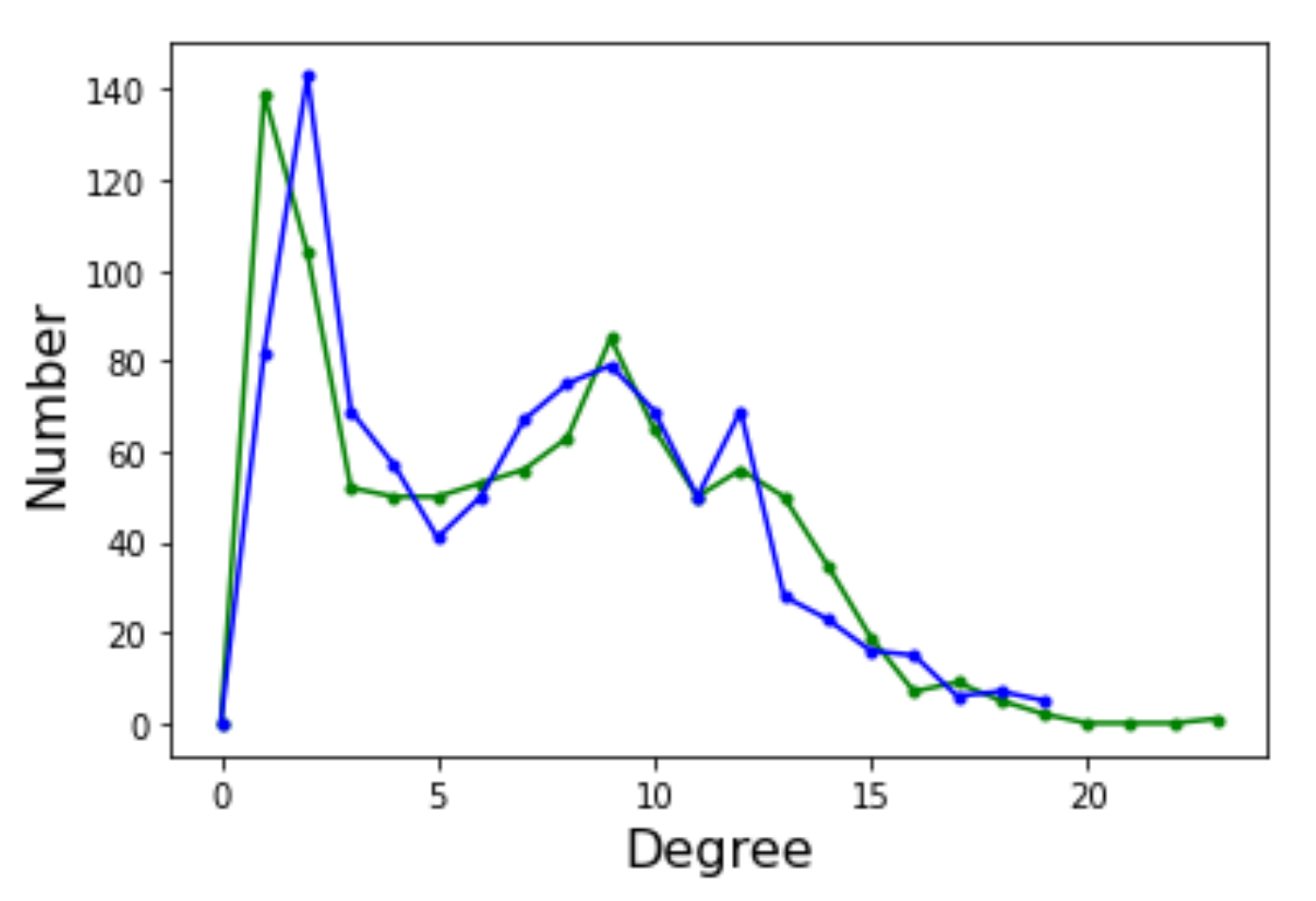}  
   \includegraphics[width=0.495\textwidth]{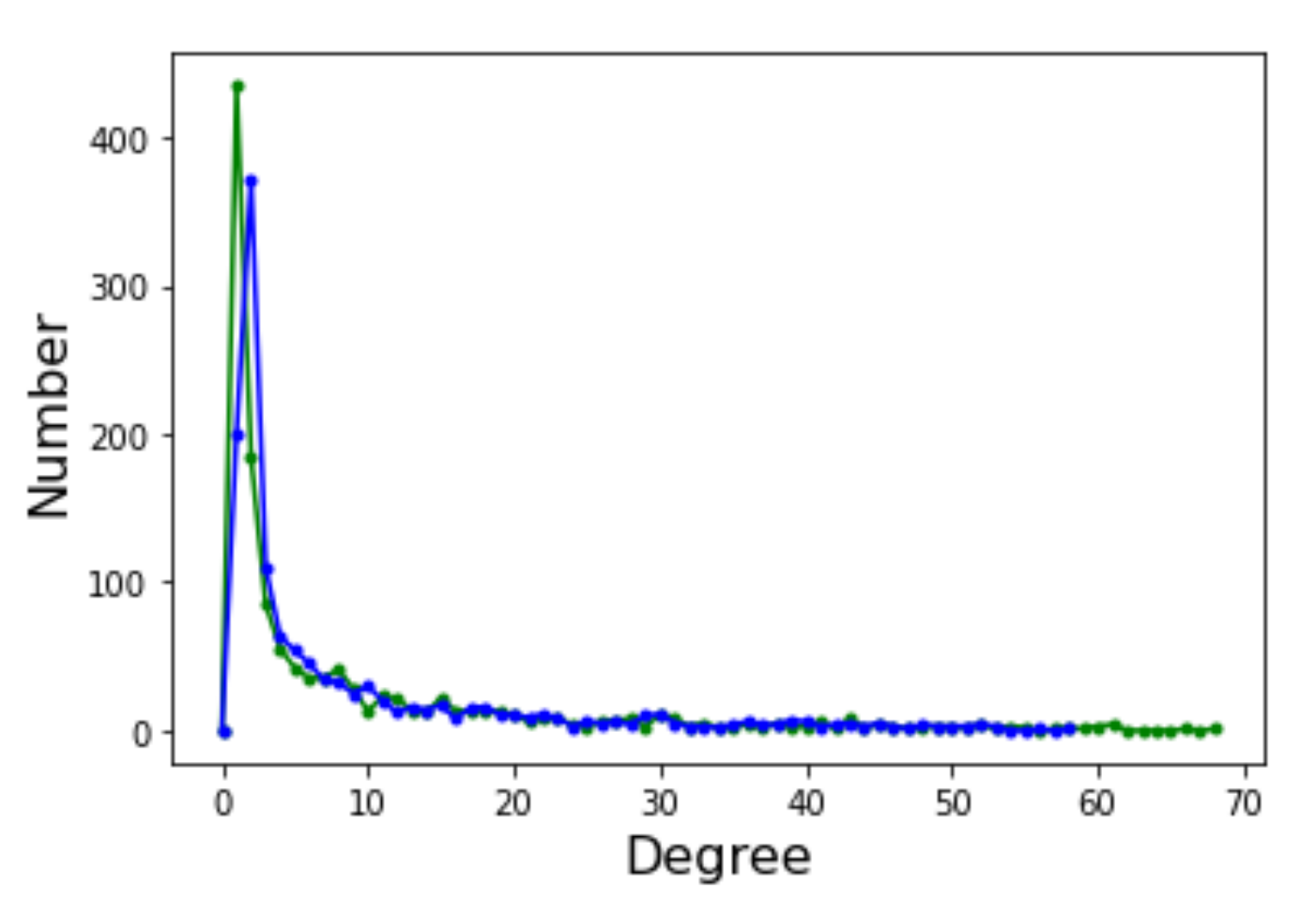}  
   \end{center}
   \vspace{-0.4cm}
  \caption{Degree distribution function for incoming (green curve) and outgoing (blue curve) edges.
  Left: Griewank function; right: Rastrigin function, both defined in two-dimensional space.}
\label{dd}
\end{figure*}

The degree distribution function (DDF) $P(k)$ of a complex network provides the probability that a randomly selected node
has degree $k$. $P(k)$ may give some useful indication about the 
general structure of the network, for example whether it belongs to a known class of distributions such
as Poissonian, exponential, or power law. Ignoring the edge weights and self-loops, here we deal with directed 
graphs and thus we have
two distinct degree distributions: the incoming edges distribution and the outgoing edges distribution.
Figure~\ref{dd} depicts the incoming (green curves) and outgoing (blue curves) for Griewank (left image) and
Rastrigin (right image) sampled local optima networks.

\begin{figure*}[h!]
  \begin{center}
   \includegraphics[width=0.45\textwidth]{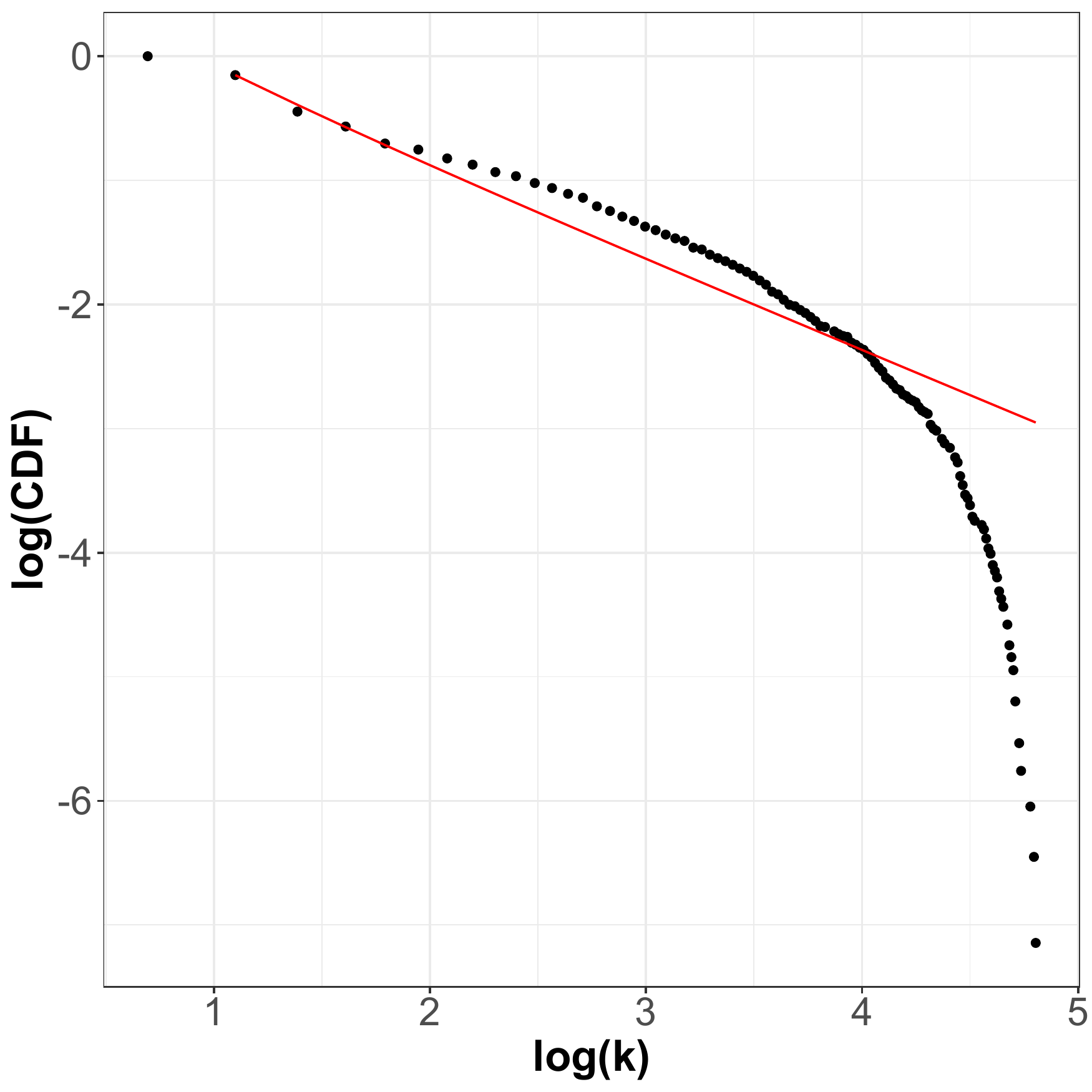}  
   \end{center}
   \vspace{-0.4cm}
  \caption{Empirical cumulative degree distribution function of all edges for the Rastrigin LON on log-log
  scale. The distribution is well fitted by a power-law shown by the line in red except near the tail cutoff, 
  as usual for finite and relatively small networks.}
\label{ll-in}
\end{figure*}

The incoming and outgoing distributions are similar in both cases but are very different between functions.
Griewank's LON gives rise to a bi-modal but otherwise rather narrow distribution while the
in and out distributions are inhomogeneous featuring longer tails for Rastrigin.
Bimodality in the Griewank's LON is due to the symmetry of the function in which each minimum has four
first neighbors, which correspond to the highest peak of the curve and 
about twelve neighbors at perturbation distance which account for the second smaller one, as can be seen in Fig.~\ref{schwefel}.
On the other hand, the long tails in the Rastrigin case indicate that there is a non-negligible 
number of LON vertices with a large number of incoming and outgoing edges, i.e., the network degree
distributions approaches a power law. This is seen by plotting the whole cumulative distribution for incoming
and outgoing edges together on double
logarithmic scale in Fig.~\ref{ll-in} with a straight line fit (exponent $1.723$, $p$-value=0.663).  This kind of distribution suggests that the high degree nodes
could play a role in dynamical processes on LONs and, in fact, we can trace a useful analogy to well known 
network robustness concepts. 

Since the work of Albert et al.~\cite{albert2000} we know
that scale-free networks are more resistant than other network topologies to random attacks to their nodes. 
Even the removal of a sizeable fraction of the nodes (and the corresponding edges) does not impact the network
function in an essential way up to a point where the network starts to fall apart. Now, and the following is
to be understood only in a metaphoric sense, not a real one, if we think of a search algorithm
as a process that jumps from node to node in the LON, we see that the Rastrigin LON will be more robust in this
sense and even if a fraction of the nodes disappear, there will remain other paths to the global optimum. In the 
Griewank case, on the other hand, the absence of highly connected nodes (hubs) will cause more damage and
the search becomes more difficult. In conclusion, there will be more paths of shorter length to the global optimum
in the Rastrigin case, making the search easier. This intuition will be confirmed in the next section by
inspecting the average path lengths.

\subsection{Paths and Distances}
\label{dist}

Using again the metaphor according to which searching the function space translates into walking through
the corresponding LON, the statistics characterization of the paths in the latter could be useful to understand the
difficulty of the search process. To start with, the average shortest path length (see Sect.~\ref{metrics}) is
$5.247$ for the Griewank LON and $5.905$ for the Rastrigin network. Since the number of nodes $N=|V|$
is $951$ for the Griewank network and $1267$ for Rastrigin, both networks are of the small world type as
$N = O (\log N)$. Moreover, the Rastrigin LON having a larger size, it is comparatively smaller, which is in
agreement with a well known result on scale-free graphs~\cite{newman2018}.
The \textit{diameter}, which is the largest among
all the shortest paths between pair of nodes, is $29$ for the Griewank LON and $34$ for the Rastrigin LON.
This is the number of hops that a searcher must perform between minima in the worst case, assuming
that minima are not visited again once found. Clearly, this is not the case for most optimization metaheuristics
in common use and, in addition, the searcher has no enough global knowledge to strictly follow the shortest 
path. Thus, these values are an interesting indication but cannot be taken too literally.

\begin{table}[h!]
\begin{center}
\small
\vspace{0.3cm}
\begin{tabular}{|l|c|c|c|}
%\hline
%\multicolumn{4} {|c|}{Paths and Diameters}  \\
\hline
 & Diameter & Av. shortest path & Av. path to GO \\
\hline
Griewank &  \;\; $26.423$  \;\;  &  \;\;\ $5.247$ & \;\; $3.70$ \;\;\\
\hline
Rastrigin &  \;\;    $33.333$ \;\;  &  \;\; $5.905$ & \;\;   $2.93$ \;\; \\
\hline
\end{tabular}
\vspace{0.3cm}
\caption{Typical distances in the Griewank network (top line) and the Rastrigin network (bottom line).
First column: diameter (largest among all shortest paths). Second column: average shortest path computed according
to eq.~\ref{av-pl} in Sect.~\ref{metrics}. Third column: average path length to the global optimum from
all other minima. }
\label{SP}
\end{center}
\end{table}
\normalsize

Another useful network measure is the average length of the shortest paths
from all the local optima to the global one for this provides an idea of the difficulty of reaching the
global optimum starting anywhere in the search region.
Using Dijkstra's algorithm, for the Rastrigin LON this quantity is equal to $2.93$; for the 
Griewank network it is $3.70$. Given that
the Rastrigin LON is larger than the Griewank LON, this shows that, on the average, a search is likely to reach
the global optimum quicker in the Rastrigin search space. This confirms from a different network point of
view the fact that Rastrigin is an  easier function than Griewank in two dimensions. Of course, this is only a qualitative
indication: a search algorithm, being unaware of the structure of the fitness landscape, could do many things other 
than traveling to the global optimum through a short path, such as cycling among local optima, and thus this information
must be taken with a grain of salt.
For the sake of clarity, the above measures are grouped in Table~\ref{SP}.

\subsection{PageRank Centrality}
\label{prc}

\begin{figure*}[h!]
  \begin{center}
   \includegraphics[width=0.495\textwidth]{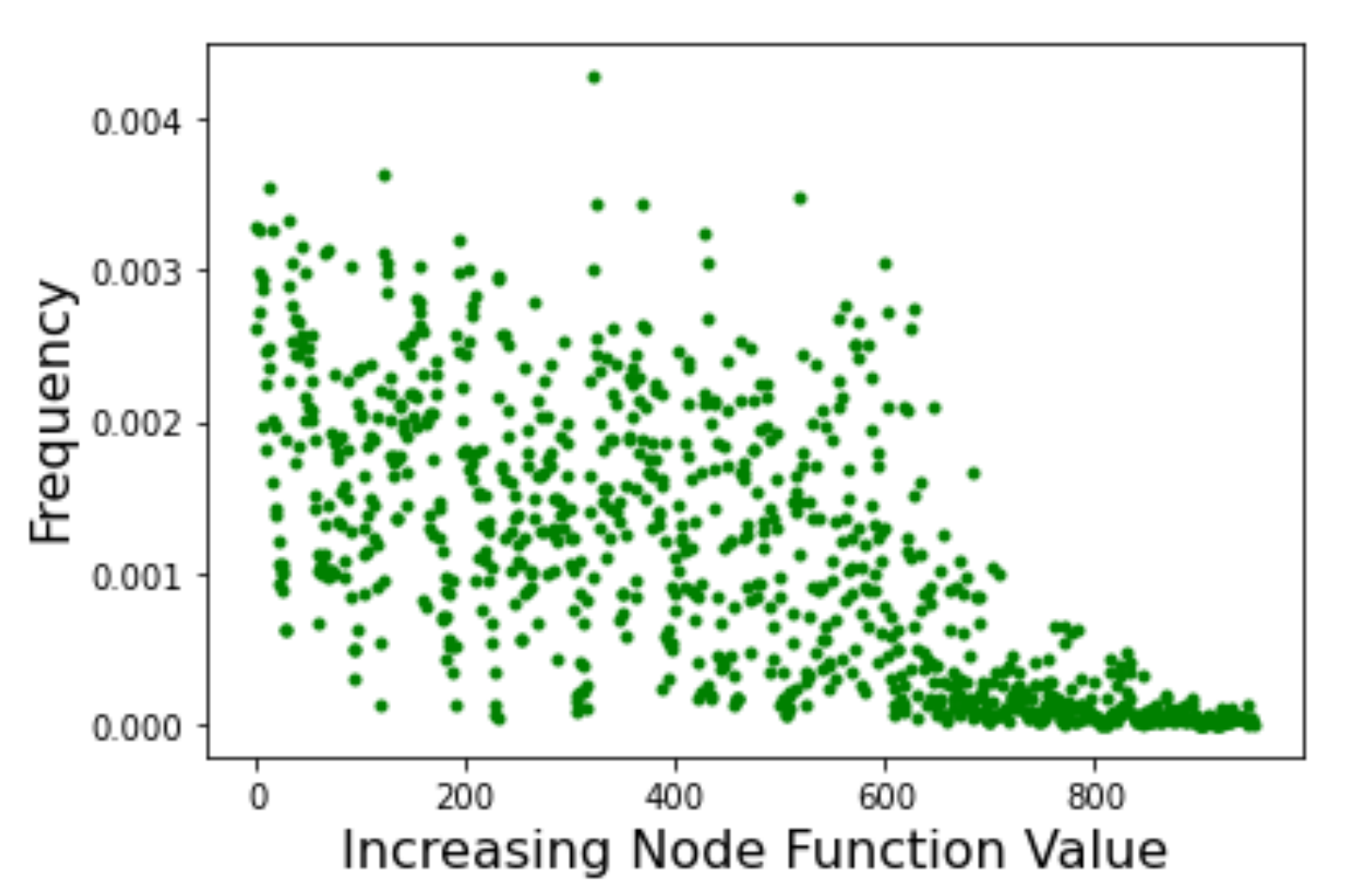}  
     \includegraphics[width=0.495\textwidth]{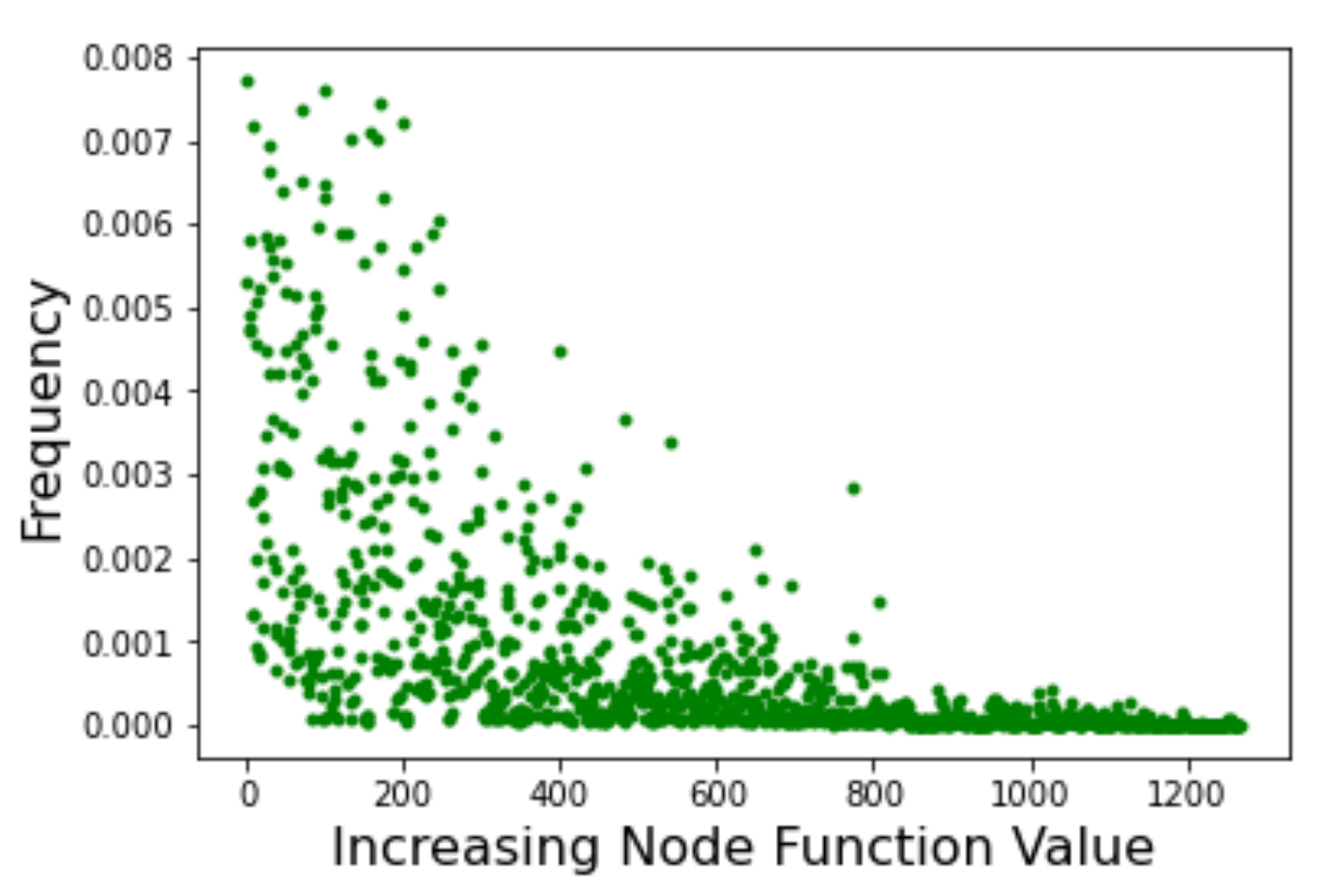} 
   \end{center}
  \caption{PageRank centrality of all minima versus their objective function values as measured by the frequency of visits 
  during a random walk. Node function values increase to the right of the x-axis. Left picture: Griewank network;
   right picture: Rastrigin network.}
\label{pr}
\end{figure*}

PageRank~\cite{PR} is an algorithm that provides a measure of the importance of a web page based
on how much it is referenced by other important pages. The aim is to capture the 
prestige of each node in order to rank pages by importance and thus reduce the amount of information
to process in a web search. The algorithm 
can be described as a random walk on the web considered as a weighted directed graph. 
In the long time limit, if some conditions are satisfied, the probability distribution reaches an invariant value. 
This equilibrium probability distribution  of the above Markov chain
gives the asymptotic frequency of visits to each node of the network.
Fig.~\ref{pr} shows the results of running PageRank on the Griewank LON (left image) and the Rastrigin LON
(right image). Graph nodes are sorted by increasing objective function value on the
x-axis, and the corresponding PageRank centrality, i.e.,  the long-term frequency
of visits of each node, is reported on the y-axis.
The results fully confirm what was found above by examining the strength of the incoming edges to nodes
 because the visit frequency depends
 on the incoming strength. Thus, in the Rastrigin LON the best minima are also the most central nodes in the
 network while many nodes in a wide range of fitness have similar centralities in the Griewank network.
 Therefore, the result in terms of PageRank is fully coherent with the picture that emerged previously
 and our qualitative argument carries over, indicating 
 that the Griewank function should be more difficult to optimize.

\subsection{Funnels and function difficulty}

\begin{figure*}[h!]
  \begin{center}
   \includegraphics[width=0.32\textwidth]{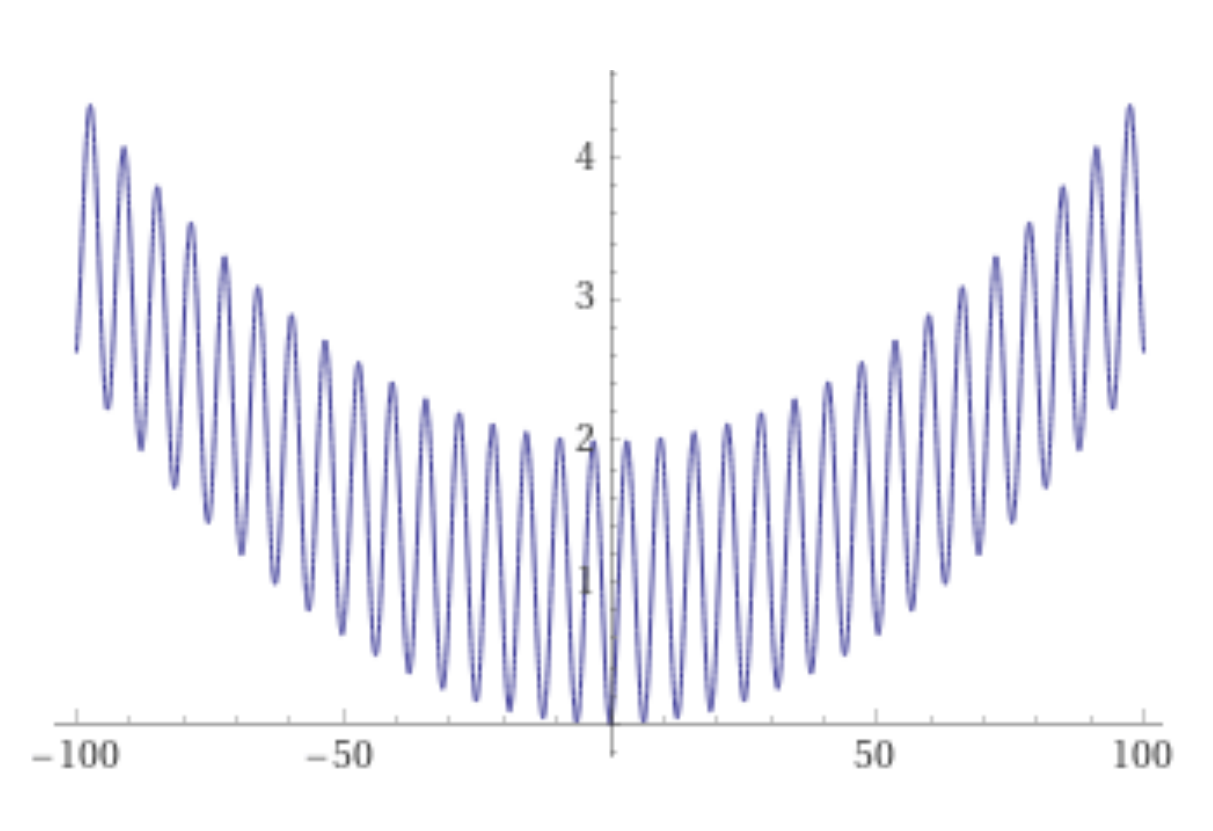}  
     \includegraphics[width=0.32\textwidth]{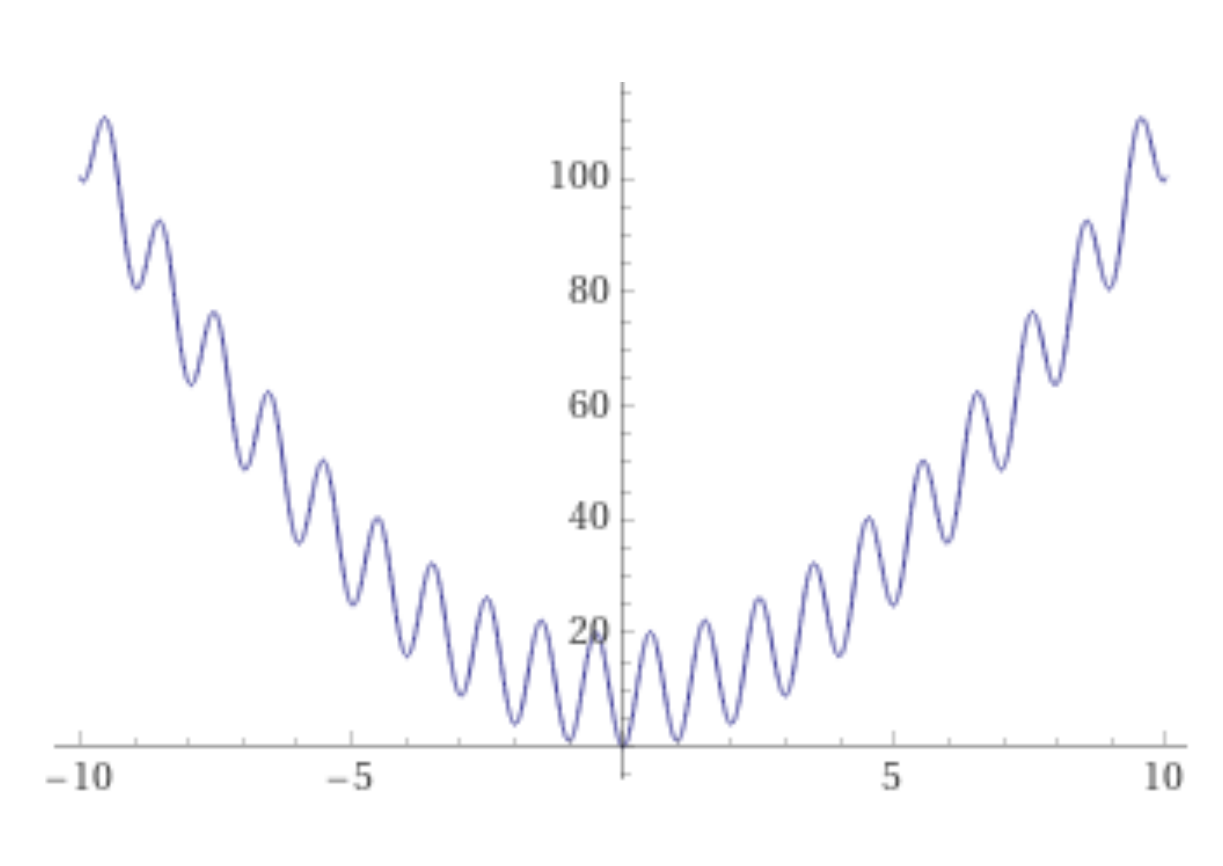} 
       \includegraphics[width=0.32\textwidth]{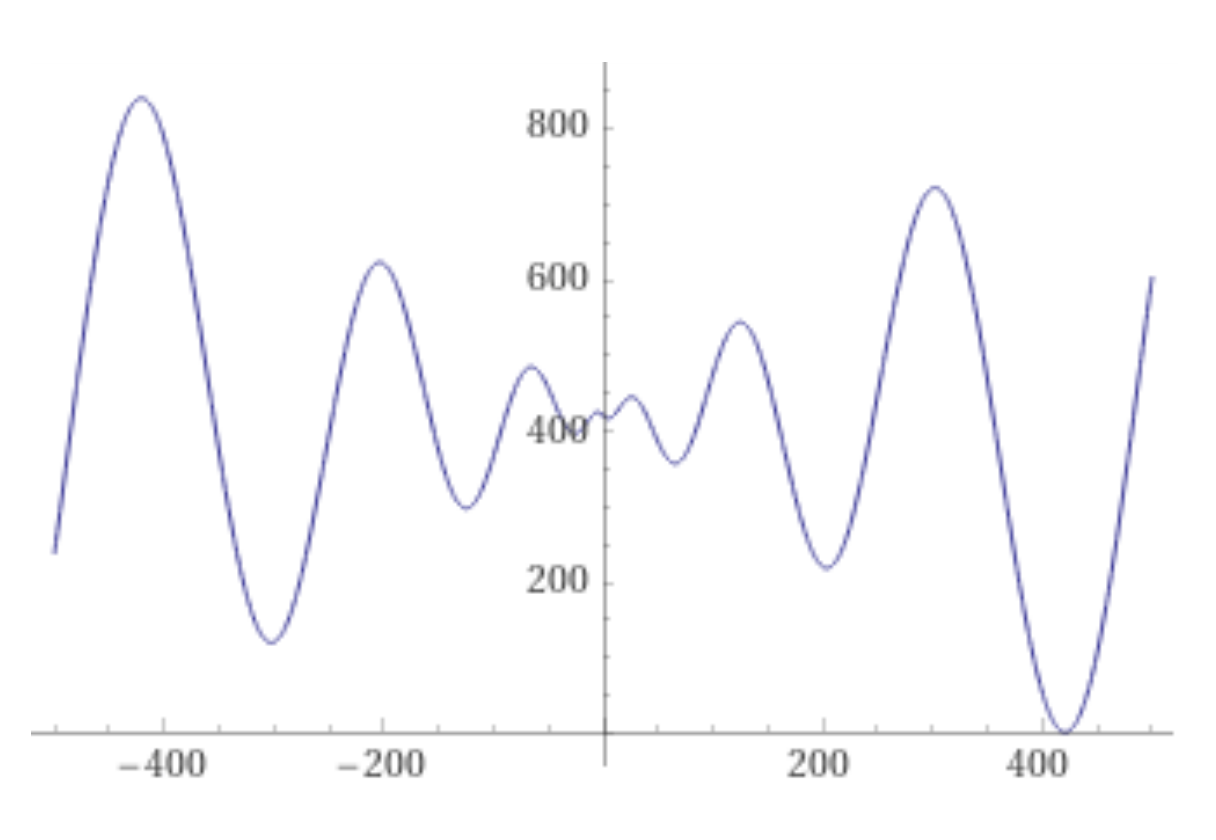} 
   \end{center}
  \caption{From left to right: one-dimensional Griewank, Rastrigin, and Schwefel functions showing a single funnel
  lanscape (Griewank, Rastrigin), and a double funnel (Schwefel). Note the different axes scales.}
\label{funnels}
\end{figure*}

The concept of a \textit{funnel} is a useful one when speaking of minimization of functions. It has been first suggested
in chemical physics to help explain the folding of natural proteins into their low-energy state through a collection
of convergent pathways in which the energy decreases systematically near the target 
structure~\cite{funnel1992,walesBook}.
Funnels, understood as a monotonic decreasing sequence of local minima that end up into the funnel bottom, called a sink, have been
analogously introduced in LONs of combinatorial spaces by Ochoa and Veerapen~\cite{ochoa2018TSP}.
Under the funnel view, fitness landscapes can be classified into single-funnel and multi-funnel. For example, and using
one-dimensional functions for clarity, Fig.~\ref{funnels} shows (from left to right) that the Griewank and the  Rastrigin functions are of
the single-funnel type, while the Schwefel function has a double funnel.  In general, multi-funnel functions are more difficult
to optimize because the search can easily get trapped into a sub-optimal funnel, but the extreme difficulty of minimizing single-funnel
energy hypersurfaces for medium to large size proteins~\cite{walesBook} suggests that this is not the only source of hardness.

The funnel structure of the LONs of some of the above test functions has been studied using non-increasing
BH sampling in~\cite{contreras2020synthetic}. The analysis shows that the funnel structure present in
the landscapes sketched in Fig.~\ref{funnels} carries over to the corresponding LON, which is obvious
for these cases but can be useful for real-world functions that have an unknown shape in parameter space.
For further details we refer the reader to~\cite{contreras2020synthetic}.

\begin{figure*}[h!]
  \begin{center}
   \includegraphics[width=0.65\textwidth]{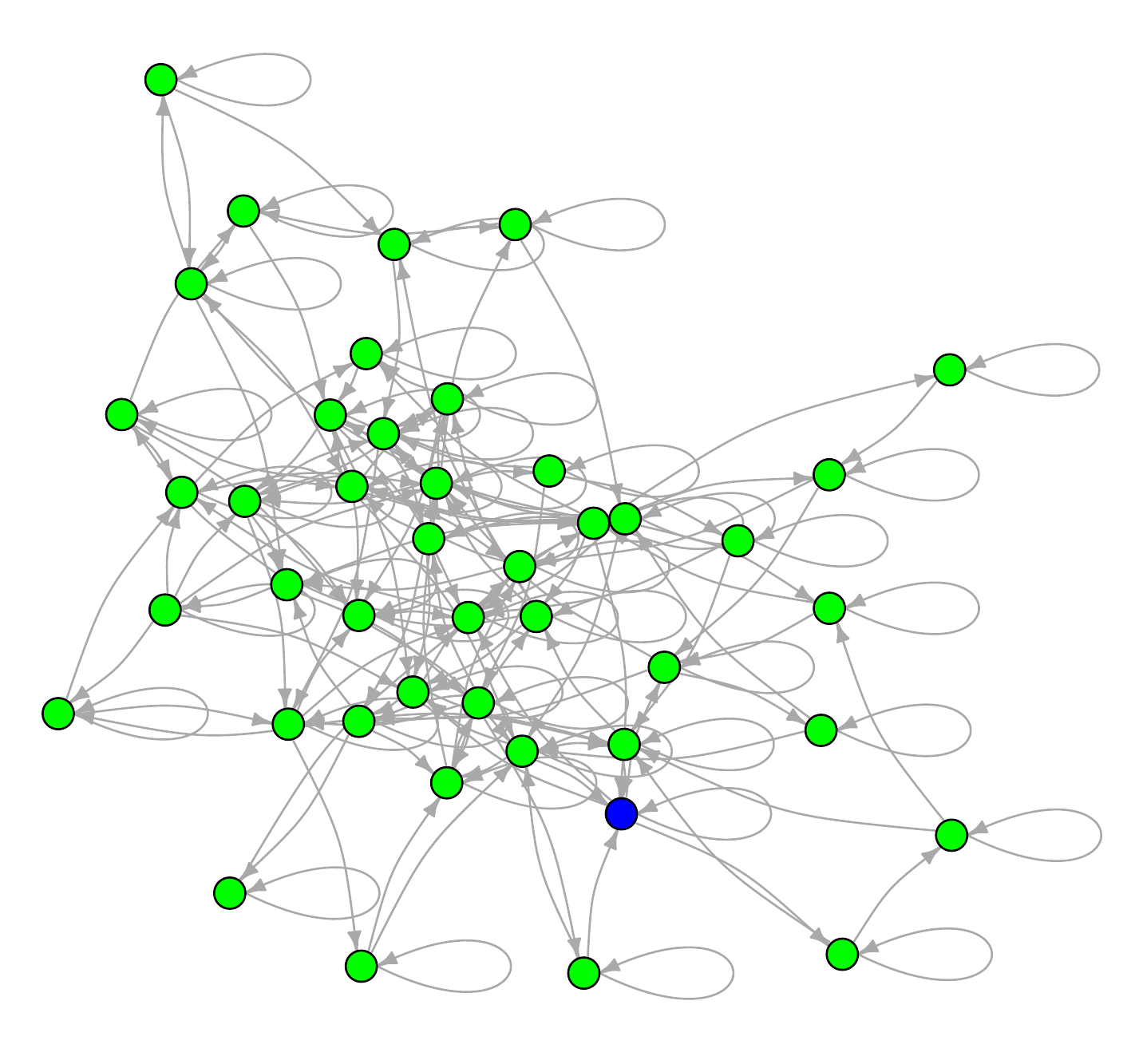}  
   \end{center}
  \caption{The sampled LON of the Schwefel function in two dimensions. Edge weights are not shown for the 
  sake of clarity. The node corresponding to the global minimum is in blue.}
\label{schwefel-lon}
\end{figure*}

To examine the LON of a multimodal and multifunnel function, we now show the sampled graph of the
two-dimensional Schwefel function defined in Sect.~\ref{functs} and whose contour lines are shown in
Fig.~\ref{schwefel}. This function is interesting for
our purposes because its global minimum is far from the origin and because it has a smaller
number of minima than either Griewank or Rastrigin. Since there are fewer minima the LON is small
($|V|=44, |E|=183$ including self-loops)
and can be easily vizualized in Fig.~\ref{schwefel-lon} where the global minimum is colored blue. Most of the
minima in the search region have been sampled but some transitions may be absent. These could be obtained
by further increasing the number of BH sampling iterations but the LON as it stands is already highly representative
of the function landscape structure.

The mean degree ignoring self-loops and edge directions is $6.32$ and the in and out degree distributions are narrow 
and centered around the mean (we do not plot them because there are too few data for meaningful
statistics). As for the typical graph distances, the average path length is $3.16596$ and the diameter
is $7$. Finally, the weighted mean path length to the global optimum from all the other minima is $2.98$, i.e., less
than three hops, from the point of view of the LON graph topology, 
remembering that if an edge has weight $w_{ij}$ then its effective length is $1/w_{ij}$ since the $w_{ij}$
represent frequency of transitions between pairs of minima; i.e., remembering that Dijkstra's algorithm
uses weights as lengths, minima connected by an edge having a
large weight are effectively ``shorter'' and thus are more likely to communicate.

Table~\ref{schwT} reports the success rate on the Schwefel function for
dimensions two, five, and ten respectively. Only the figures pertaining to Differential Evolution have been
reported as BH has troubles converging to the global minimum for this function within the allowed budget of
function evaluations.

\begin{table}[h!]
\begin{center}
\vspace{0.3cm}
\small
\begin{tabular}{|l|c|c|c|c|c|c|}
\hline
\multicolumn{7} {|c|}{Success Rates}   \\
\hline
% aaa & bbb& \multicolumn{2}{c|}{fusion1} & \\
Schwefel & \multicolumn{2}{c|}{n = 2} & \multicolumn{2}{c|}{n = 5} & \multicolumn{2}{c|}{n = 10} \\
\hline
& \multicolumn{2}{c|}{$0.910\;(1115)$} & \multicolumn{2}{c|}{$0.866\;(6265)$} & \multicolumn{2}{c|}{$0.675\;(26992)$} \\
\hline
%& $DE$ & $BH$ &  $DE$  & $BH$ & $DE$  & $BH$ \\
%\hline
%\hline
%&  \;\; $0.910\;(1115)$  \;\;  &  \;\;\ $$  \;\;& \;\; $0.866\;(6265)$ \;\;&  \;\;$$ \;\; &  \;\;$0.675\;(26992)$ \;\; &  \;\;$$ \;\;  \\
\end{tabular}
\vspace{0.3cm}
\caption{Fraction of optimization runs that found the global minimum for the Schwefel
function using DE for dimensions $n=2,5,10$.
The averages are over $500$ optimization runs in each case. The search region is $[-500,500]^n$.
The allocated budget of function evaluations is $3.0 \times 10^3, 10.0 \times 10^3$, and
$30.0 \times 10^3$ for $n=2,5$ and $10$ respectively.
The mean number of function evaluations
when the global minimum has been found is in parentheses.}
\label{schwT}
\end{center}
\end{table}

Comparing now Schwefel in two dimensions with the corresponding results using DE 
for Griewank and Rastrigin (see Table~\ref{hard}) it is apparent that Schwefel is far easier than Griewank and
approximately as difficult as Rastrigin, in spite of the higher number of local minima of the latter. We see
thus that Schwefel's multifunnel is more than compensated by the non-separability and high multimodality
of the single funnel Griewank. In conclusion, multi-funnels surely make the search harder but there are
several other factors influencing optimization difficulty such as high multimodality, non-separability, deception and
conditioning, some of them being of numerical nature rather that topological.  This is why modern 
benchmark test suites try to incorporate all of these features in order to duplicate the
typical difficulties that are believed to occur in continuous domain search in practice. A good example
of this approach is reference~\cite{hansen2020coco}.

\subsection{Scaling to higher dimension}
\label{scaling}

In this section we investigate how and when the indications drawn in the two-dimensional case extend
 to higher dimension. First of all, it is useful to get an understanding of how the number of local 
 minima scales with dimension for the multimodal functions used in the text. To this end, a reasonable estimate, which is a lower bound, can be obtained
 with a simple multistart algorithm as described in pseudocode~\ref{multi}. The algorithm
 generates a large number $q$ of random starting points and then, for each of them, locally minimizes the function. 
  
\begin{algorithm}
\caption{Multistart Minima Sampling}
\label{multi}
\begin{algorithmic}[0]
\Require $f(.)$, bounding box $B = x \in[a,b]^n$, number of starting points $q$
\State create empty list of minima $M$
\For{$i  \leftarrow 1$ to $q$}
\State $s \leftarrow$ generate a random solution in $B$
%\IF{$f(y) < f(x)$}
\State$m \leftarrow $ minimize(f(s))
\If {$m  \not\in M$ and $m$ not out of bounds $B$}
     \State add $m$ and $f(m)$ to $M$
\EndIf
\EndFor
\State \textbf{return} $M$
\end{algorithmic}
\end{algorithm}
 
 The results obtained using this technique are summarized  in Table~\ref{sizes} where the number of local 
 optima is reported for dimensions $n=2,5$, and $ 10$ for each function. 
 The hyperboxes considered for each function are $\textbf{x} \in[-60,60]^n$ for the Griewank function, 
$\textbf{x} \in[-30,30]^n$ for Rastrigin, and $\textbf{x} \in[-500,500]^n$ for the Schwefel function.
 Except for the case $n=2$ for which
  the figures are close to the actual values, these numbers are
 lower bounds and thus the true number of optima is surely underestimated. 
 
 \begin{table}[h!]
\begin{center}
\vspace{0.3cm}
\small
\begin{tabular}{|l|c|c|c|c|c|c|}
\hline
\multicolumn{7} {|c|}{Number of Local Minima}   \\
\hline
% aaa & bbb& \multicolumn{2}{c|}{fusion1} & \\
& \multicolumn{2}{c|}{n = 2} & \multicolumn{2}{c|}{n = 5} & \multicolumn{2}{c|}{n = 10} \\
\hline
Schwefel & \multicolumn{2}{c|}{$\sim 50$} & \multicolumn{2}{c|}{$\sim 11000$} & \multicolumn{2}{c|}{$> 85000$} \\
\hline
Rastrigin & \multicolumn{2}{c|}{$\sim 3600$} & \multicolumn{2}{c|}{$> 100000$} & \multicolumn{2}{c|}{$> 100000$} \\
\hline
Griewank & \multicolumn{2}{c|}{$\sim 528
$} & \multicolumn{2}{c|}{$> 95000$} & \multicolumn{2}{c|}{$> 60000$} \\
%& $DE$ & $BH$ &  $DE$  & $BH$ & $DE$  & $BH$ \\
\hline
%\hline
%&  \;\; $0.910\;(1115)$  \;\;  &  \;\;\ $$  \;\;& \;\; $0.866\;(6265)$ \;\;&  \;\;$$ \;\; &  \;\;$0.675\;(26992)$ \;\; &  \;\;$$ \;\;  \\
\end{tabular}
\vspace{0.3cm}
\caption{Estimated number of local minima for Schwefel, Rastrigin, and Griewank functions for dimensions
$n=2,5,10$ defined
in $\textbf{x} \in[-500,500]^n$, $\textbf{x} \in[-30,30]^n$, and $\textbf{x} \in[-60,60]^n$ respectively.}
\label{sizes}
\end{center}
\end{table}

The values in Table~\ref{sizes} are a manifestation of what is usually called the ``curse of dimensionality" meaning
the rapid increase of problem volume associated with increasing space dimension. For the highly multimodal
functions we consider here the number of minima consequently grows rapidly, making sampling, and also global optimization, more difficult and time consuming for higher dimension. We mention in passing that
this is not true for the Griewank function in which the number of minima first increases but then decreases 
for $n > 5$ for the reasons sketched in Sect.~\ref{perfs}.
For the Schwefel function going to $n=5$ is not a problem but for Griewank and especially Rastrigin the
process, although it can technically be performed, is very lengthy if good sampling is required. 

\begin{figure*}[h!]
  \begin{center}
   \includegraphics[width=0.7\textwidth]{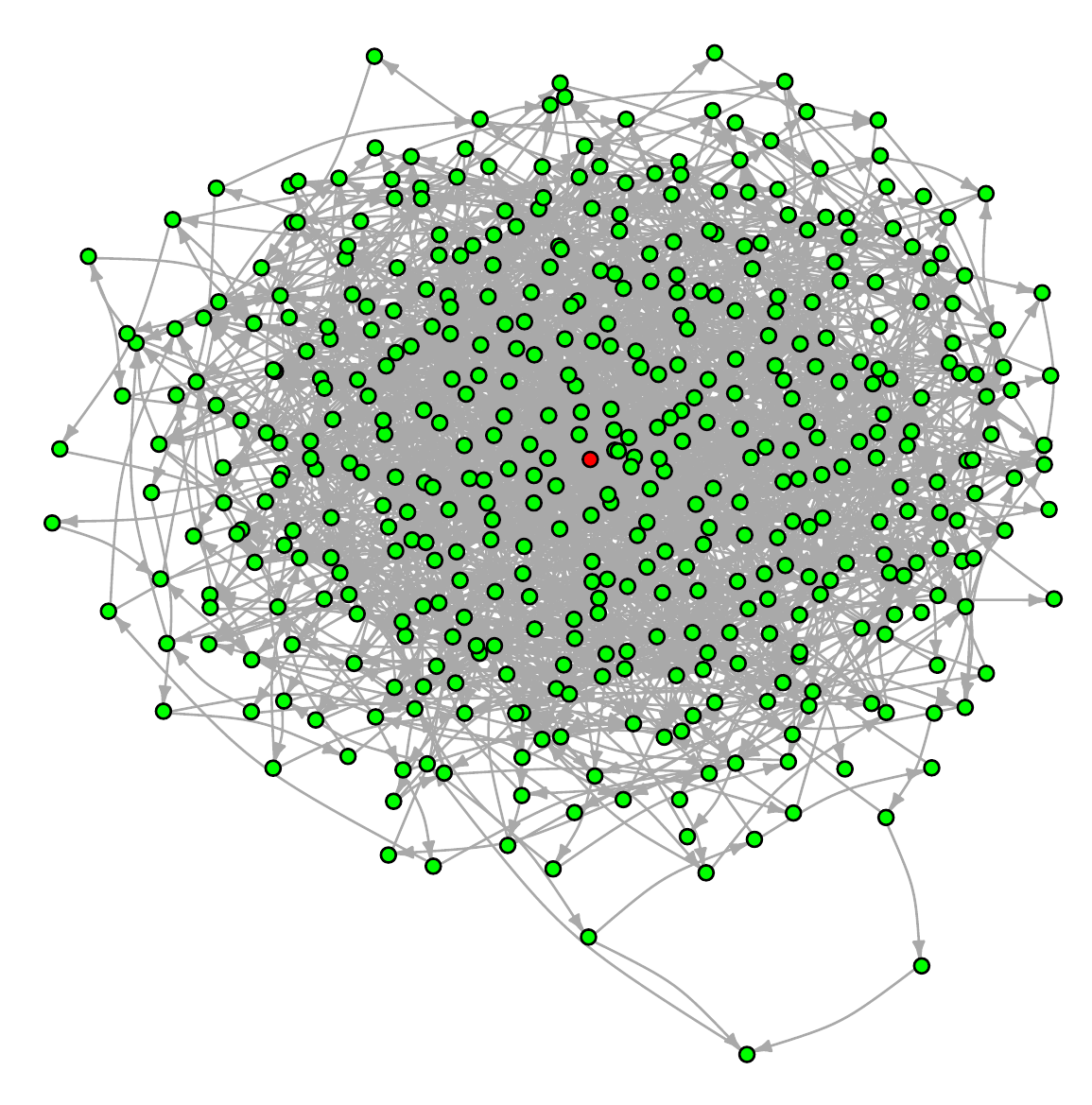}  
   \end{center}
  \caption{The LON of the Griewank function in five dimensions in the box $[-10,10]^5$. Self-loops, edge weights and node labels are not shown for the  sake of clarity. The node corresponding to the global minimum is in red.}
\label{GW5-lon}
\end{figure*}

For the sake
of illustration here we will take the Griewank function at $n=5$ in the reduced box $\textbf{x} \in[-10,10]^5$
for which algorithm~\ref{multi} finds about $500$ local minima. The resulting LON is shown in Fig.~\ref{GW5-lon}.
 
\begin{figure*}[h!]
  \begin{center}
   \includegraphics[width=0.58\textwidth]{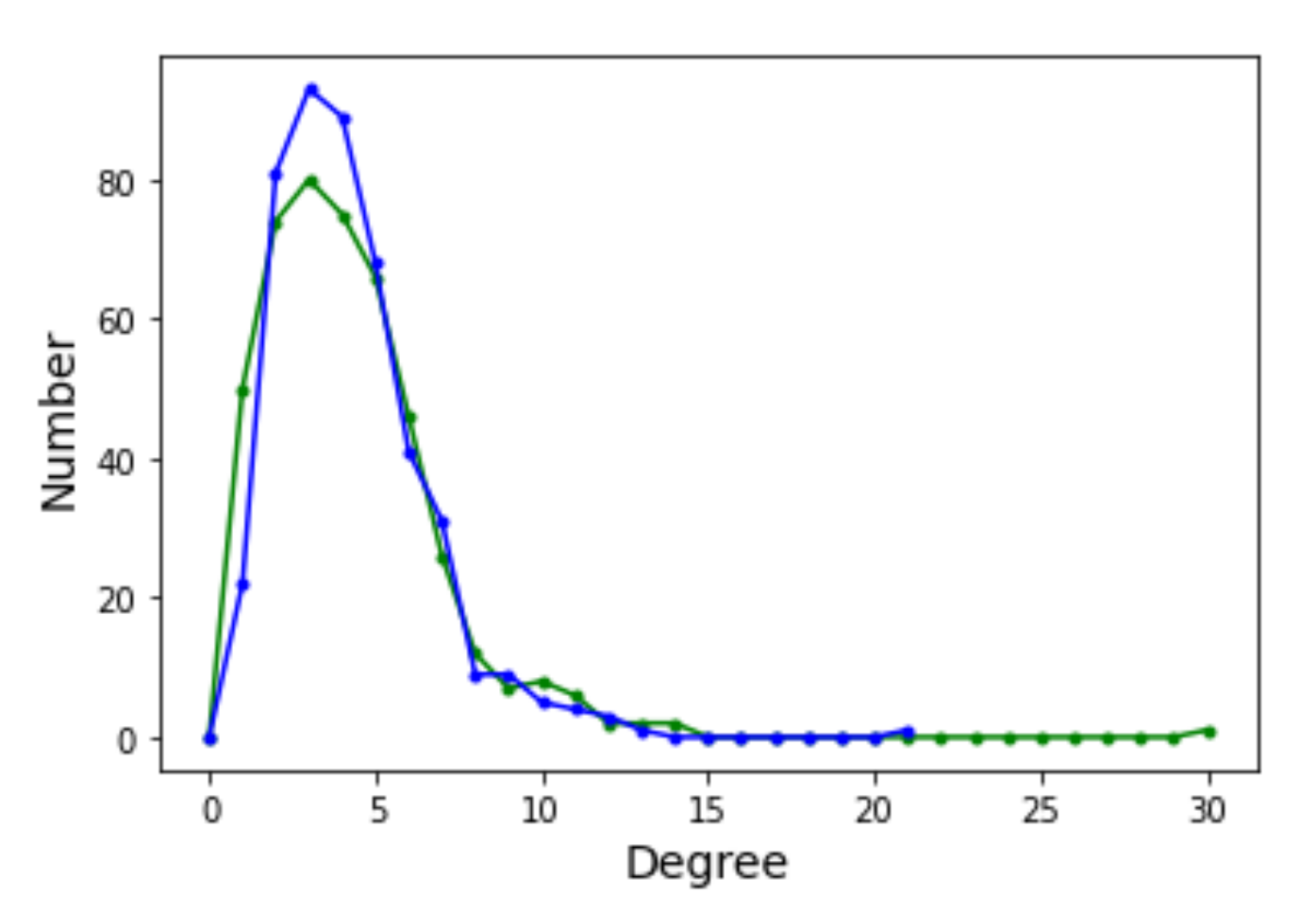}  
   \end{center}
  \caption{Degree distribution functions for the Griewank LON in five dimensions.
  Green curve: incoming edges; blue curve: outgoing edges.}
\label{ddf-gw-5}
\end{figure*}

The graph has $457$ vertices and $1939$ edges without considering self-loops for a mean degree of $8.49$.
The degree distribution function is shown in Fig.~\ref{ddf-gw-5}. With respect to the DDF for the case
$n=2$ (see Fig.~\ref{dd} in Sect.~\ref{DDF}) we observe that the shape of the curves are intermediate
between Griewank and Rastrigin in two dimensions. In the present case with $n=5$ the curves have a longer
tail to the right but not as extended as for Rastrigin. This suggests that there are more nodes with
relatively high in and out degree and we have seen in Sect.~\ref{DDF} that this is an indication that
searching the optima becomes easier. Indeed, the Griewank function in five dimensions is less hard
than the same function in two dimensions according to Tables~\ref{hard} and~\ref{D5-10} in Sect~\ref{perfs}.

\begin{figure*}[h!]
  \begin{center}
   \includegraphics[width=0.485\textwidth]{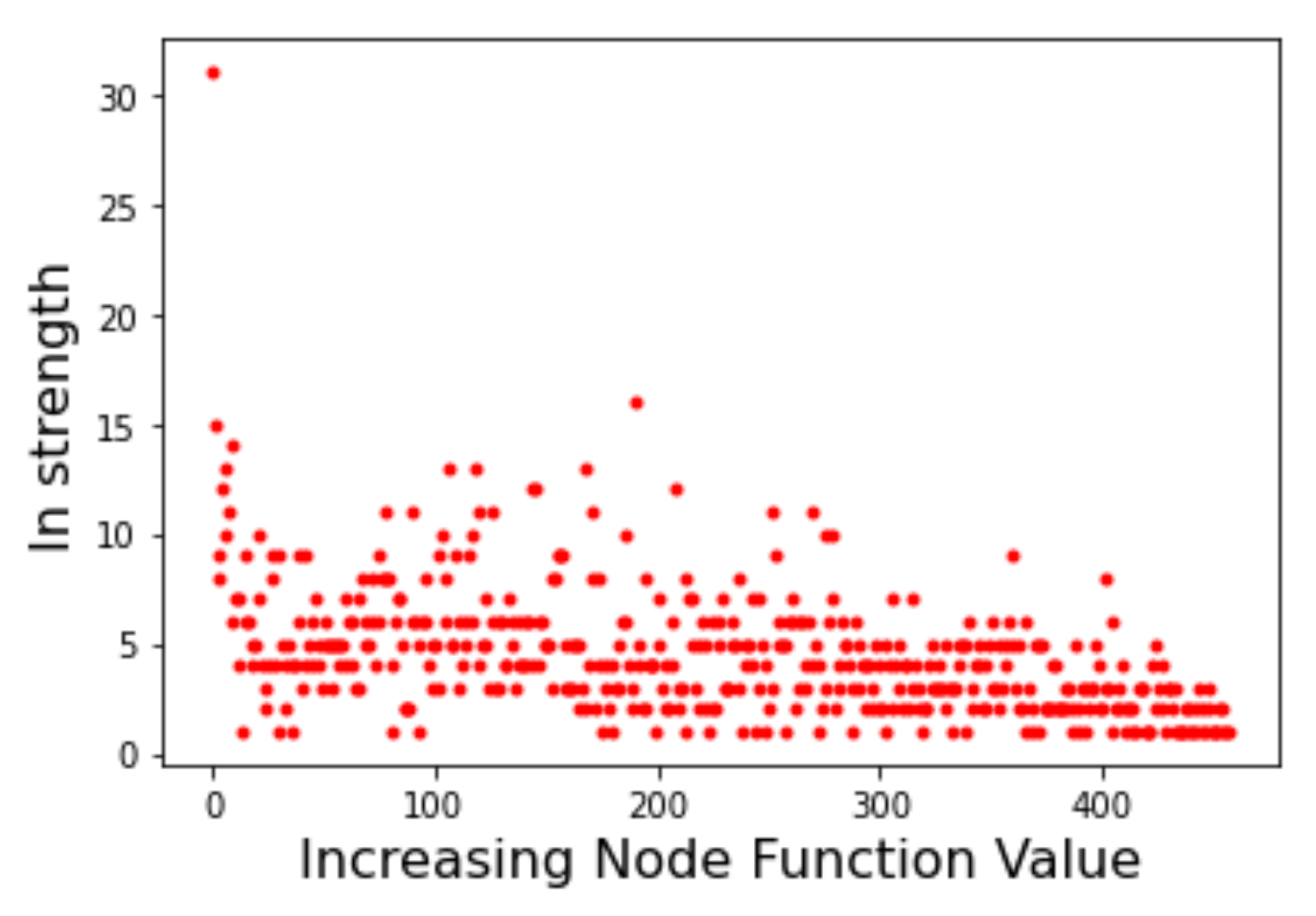}  
     \includegraphics[width=0.5\textwidth]{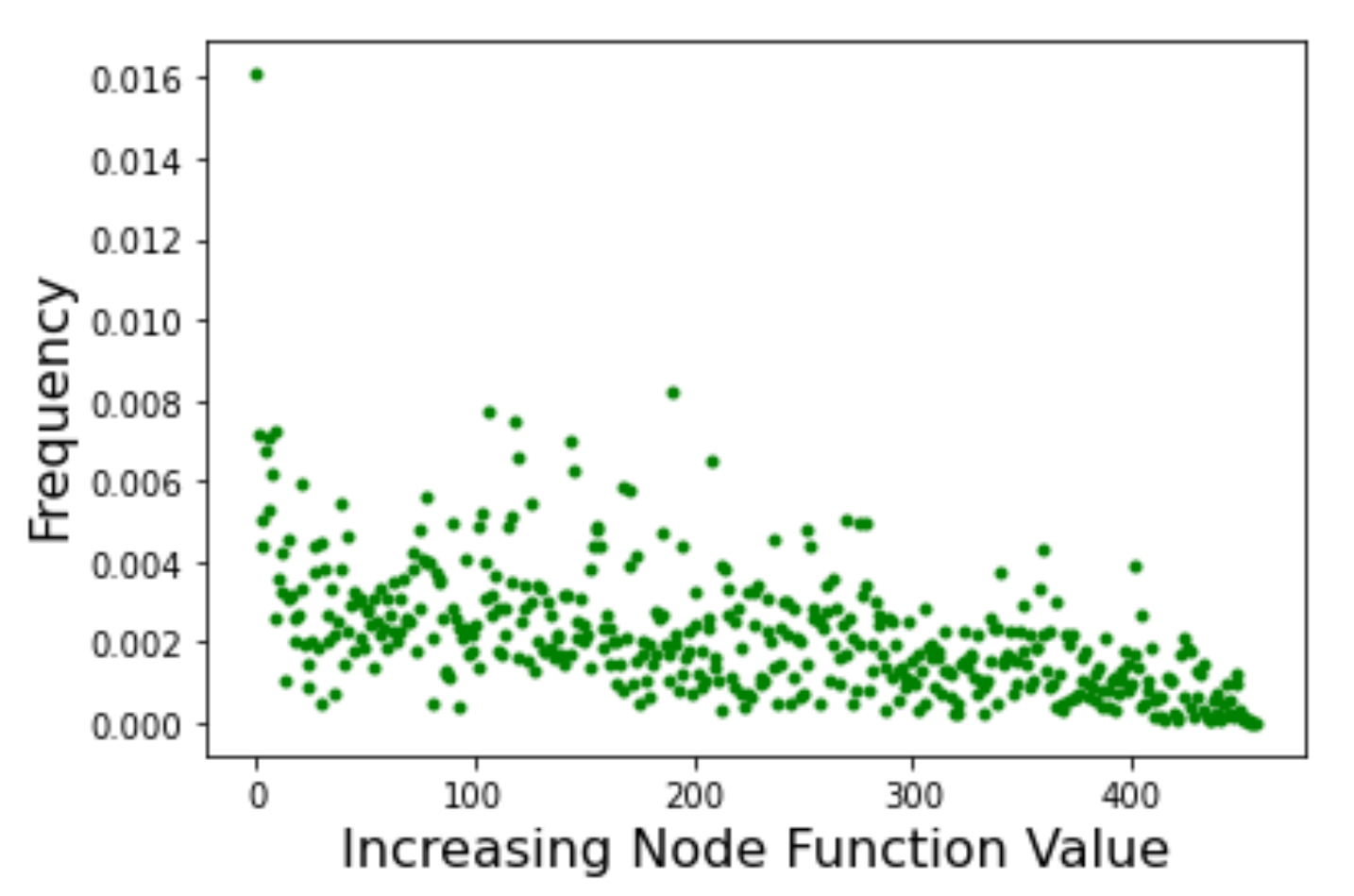}  
   \end{center}
  \caption{Griewank function at dimension $n=5$. Nodes are sorted by increasing objective function value.
  Left image: incoming strength.
  Right image: node centrality according to the Page Rank algorithm.}
\label{PR5}
\end{figure*}

The above indications are confirmed by the incoming edges strength and by the related Page Rank
centrality depicted in Fig.~\ref{PR5} with lower function values to the left. Here we see that 
high-fitness nodes tend to have, on average, a higher
incoming strength and, correspondingly, high centrality. Indeed, the global optimum is the node with
both highest strength and highest centrality.

The average path length is $4.28$, logarithmic in $|V|$, but the region of interest, i.e., $\textbf{x} \in[-10,10]^5$,
is smaller than the one considered at $n=2$ which was $\textbf{x} \in[-60,60]^2$. In fact, the shrinking
of the box is approximately compensated by the increase in volume at higher dimension.

A benefit of the LON representation can be seen here. A problem that would be very hard or even
impossible to describe in the original five-dimensional metric space is transformed into a graph
in which neighbor and other graph relationships corresponding to features in the original problem
space can be easily computed, and sometimes even visualized. This is an advantage because in high dimension
the concept of closeness becomes unclear.

\begin{figure*}[h!]
  \begin{center}
   \includegraphics[width=0.6\textwidth]{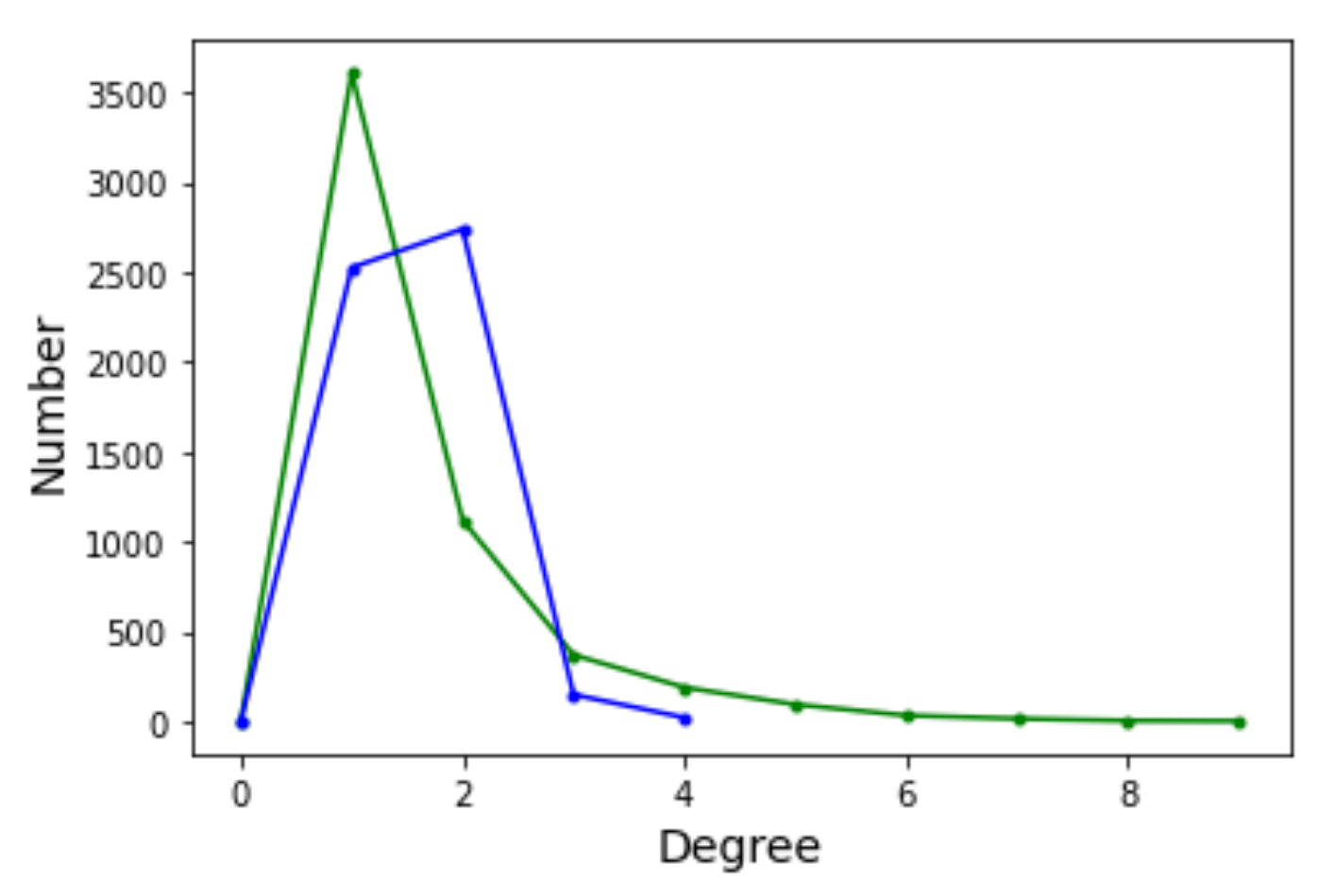}  
   \end{center}
  \caption{Degree distribution functions for the Schwefel LON in five dimensions.
  Green curve: incoming edges; blue curve: outgoing edges.}
\label{schw-DD}
\end{figure*}

As a final example, let us consider the LON corresponding to the Schwefel function in the domain
$\textbf{x} \in[-500,500]^5$. This LON is much larger with $|V| = 5436 $ and $|E| = 8536$ excluding self-loops,
which is clearly too large for direct visualization to make sense.
In this case we know from Table~\ref{sizes} that there are about $11000$ local minima. The sampling
process finds half of them in reasonable time but the number of directed edges, i.e., the transitions between minima,
is certainly underestimated with a mean degree of only $3.14$ not counting self-loops. This makes computing
path lengths less reliable and so we do not do it. The DDF is also influenced by the undersampling of graph
edges of course, but it is narrow and concentrated around the mean anyway, see Fig.~\ref{schw-DD}. 
However, the incoming edge distribution (green curve) has a longer tail, which is an indication of the presence
of some nodes with a high incoming degree. As we saw in Sect.~\ref{resul} this suggests that those nodes are
easier to reach in a search process. 

\begin{figure*}[h!]
  \begin{center}
   \includegraphics[width=0.6\textwidth]{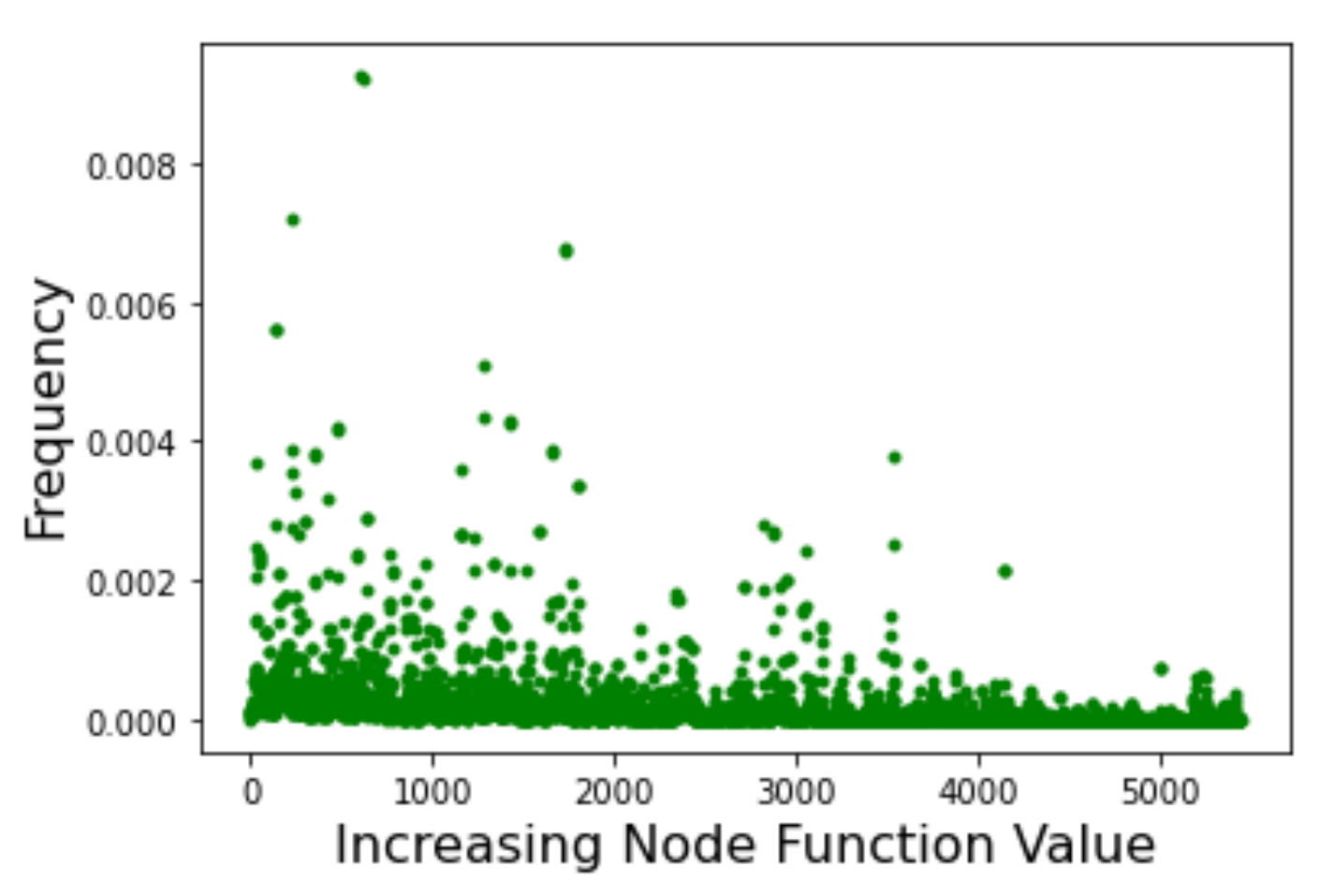}  
   \end{center}
  \caption{The Page Rank centrality of nodes in the Schwefel LON. Nodes represent function
  minima and are ordered by increasing objective function value.}
\label{schw-PR}
\end{figure*}

Below we also give the Page Rank centrality results in Fig.~\ref{schw-PR}.
This scatter plot provides qualitative confirmation that, indeed, the most central nodes are among those
corresponding to low values of the objective function of the corresponding minima. In other words, good
minima should be relatively easy to locate during search.
This is in agreement with the fact that Schwefel should remain ``easy'' in higher dimensions as 
shown in Table~\ref{schwT}.

\section{Conclusions}

In this work we have applied the LON model, which abstracts
the global fitness landscape of a continuous function in some domain of  multidimensional 
real space  into a weighted directed graph. The approach
had been applied previously with success to discrete spaces of the combinatorial optimization
type. Starting with simple two-dimensional functions chosen among multimodal standard test
functions, we showed that the methodology for sampling
the given domain can be based on basin hopping, a trajectory method that jumps from a local
minimum to another neighboring local minimum. The LON graph thus obtained can be examined using
a number of network metrics that, altogether, qualitatively correlate with the hardness
of the corresponding function, as empirically measured by success rate using some metaheuristic
for global optimization. The method scales to higher dimension to a certain extent but, as all computations on quickly increasing search volumes, it suffers from the ``curse of dimensionality'' phenomenon. Although 
sampling could be performed in principle for dimensions equal or larger than ten, in practice it takes 
too long and the graphs obtained thus
are too sparse to be a good representation of the original real space landscape. This is especially the case
for highly multimodal functions which are the more adapted and the more interesting for
 the LON representation but whose number of minima grows too quickly with increasing dimension.
 Clearly, the limitation is less stringent for other slowly varying functions.
 Nevertheless, we have shown on five-dimensional versions of the functions that the method has
 the merit of being able to represent relationships between function minima and the transitions
 between them in a way that would not be possible in the original multidimensional metric space on
 which these functions are defined. Although the present work is just a first step, examining the 
 LON of a given function or function class in various ways,
 could in principle provide useful general information for optimization methods of the metaheuristic type as
 these approximate algorithms make heavy use of the underlying search space structure. This aspect
 is left for future work but see
 Homolya and Vinko~\cite{homolya2019} for a recent study in which LONs are used in the context of
 Memetic Differential Evolution. Also,
 we have used difficult but contrived functions that are typically found in standard benchmark test suites.
 But standard test functions are not necessarily representative of functions arising in applications.
 It would thus be interesting to test the methodology on real-world continuous optimization problems
 in the future. A first step in this direction has been done in~\cite{contreras2020synthetic} and
 applications to simple machine learning problems have been shown using a related approach
  in~\cite{ballard2017ML}.

%\reftitle{References}


%\bibliography{your_external_BibTeX_file}

%=====================================
% References, variant B: internal bibliography
%=====================================

\end{document}